%% file: ams.tex
\documentclass[10pt, twocolumn, prd, showpacs, amsmath, amssymb, aps, floats, floatfix, nofootinbib]{revtex4}

\usepackage{bm}
\usepackage{array}
\usepackage{graphicx}
\usepackage{subfigure}
\usepackage{multirow}
\usepackage{dcolumn}
\usepackage[]{color}
\usepackage[CJKbookmarks=true,colorlinks,linkcolor=blue,anchorcolor=red,citecolor=green]{hyperref}
\usepackage{cleveref}

\def\kpc{\,\mathrm{kpc}}
\def\km{\,\mathrm{km}}

\def\GeV{\,\mathrm{GeV}}
\def\MeV{\,\mathrm{MeV}}
\def\GV{\,\mathrm{GV}}
\def\MV{\,\mathrm{MV}}
\def\cm{\,\mathrm{cm}}
\def\s{\,\mathrm{s}}
\def\sr{\,\mathrm{sr}}
\newcolumntype{p}{D{,}{\pm}{-1}}

\begin{document}
\title{Quantitative study of the AMS-02 electron/positron spectra:
implications for the pulsar and dark matter properties}

\author{Su-Jie Lin}
\author{Qiang Yuan}
\author{Xiao-Jun Bi}

\affiliation{Key Laboratory of Particle Astrophysics, Institute of High
Energy Physics, Chinese Academy of Science, Beijing 100049, P.R.China
}

\begin{abstract}

The AMS-02 has just published the unprecedentedly precise measurement of
the cosmic electron and positron spectra. In this paper we try to give
a quantitative study on the AMS-02 results by a global fitting to the
electron and positron spectra, together with the updated positron fraction
data. The Markov Chain Monte Carlo algorithm is adopted to do the fitting.
The primary electron spectrum and the parameters for pulsars or
dark matter which contribute extra positrons are determined simultaneously.
We find that there is a hardening of the primary electron spectrum at
$\sim 60$ GeV. With such a new feature at the background spectrum both
the pulsars and dark matter can explain the AMS-02 results very well.
The dark matter scenario shows a drop at positron fraction at $\sim 300$
GeV, however, suffers very strong constraints from Fermi $\gamma$-ray
observations. The fitting results also suggest that the propagation model
with convection may be more favored by the lepton data than the
reacceleration model.

\end{abstract}

\date{\today}

\maketitle

\section{Introduction} \label{section_introduction}

There were large progresses in the measurements of the cosmic ray (CR)
lepton fluxes in recent years. Satellite experiments such as the Payload
for Antimatter Matter Exploration and Light-nuclei Astrophysics (PAMELA) and
the Fermi Large Area Telescope (Fermi-LAT), as well as the balloon-borne
detector such as the Advanced Thin Ionization Calorimeter (ATIC) and the
ground-based Cherenkov telescopes like the High Energy Stereoscopic System
(HESS) and the Major Atmospheric Gamma-ray Imaging Cherenkov Telescopes
(MAGIC), have improved the uncertainties of the measurements from order
of magnitude down to several tens of percents \cite{Adriani:2008zr,
Adriani:2011xv,Abdo:2009zk,Ackermann:2010ij,Chang:2008aa,Aharonian:2008aa,
BorlaTridon:2011dk}. The space station experiment Alpha Magnetic Spectrometer
(AMS-02), launched in May 2011, further improve the measurement precision
of the CR fluxes by an order of magnitude due to larger exposure and
much better control of the systematics \cite{Aguilar:2013qda}.
With the AMS-02 result, we could perform the study of CRs in a more quantitative
way instead of the qualitative studies \cite{Yuan:2013eja,Feng:2013zca,
Jin:2013nta,DiMauro:2014iia}.

The most interesting features found in the CR leptons are the excess of
the positrons compared with the secondary background expectation from CR
nuclei interaction with the interstellar medium (ISM) \cite{Barwick:1997ig,
Aguilar:2007yf,Adriani:2008zr,Aguilar:2013qda}. Combining with the electron
(or total electron/positron) spectra \cite{Adriani:2011xv,Abdo:2009zk,
Chang:2008aa,Aharonian:2008aa,BorlaTridon:2011dk} implies that there
should be extra sources emitting electron-positron pairs. The proposed
models of the extra sources include the astrophysical sources such as
pulsars \cite{1970ApJ...162L.181S,2001A&A...368.1063Z,Yuksel:2008rf,
Hooper:2008kg,Profumo:2008ms},
interaction occurring around the CR acceleration sources \cite{Blasi:2009hv,
Hu:2009zzb,Fujita:2009wk}, as well as the dark matter (DM) annihilation/decay
\cite{Bergstrom:2008gr,Barger:2008su,Cirelli:2008pk,Yin:2008bs,Zhang:2008tb,
Bergstrom:2009fa}. One can refer to the reviews for detailed description
of the relevant models to explain the electron/positron excesses
\cite{He:2009ra,Fan:2010yq,Serpico:2011wg,Cirelli:2012tf,Bi:2013dwa}.

Given the data are more abundant and precise, we developed a global
fitting tool which employs a Markov Chain Monte Carlo (MCMC,
\cite{Lewis:2002ah}) method to sample the high-dimensional parameter space 
of the CR propagation and injection \cite{Liu:2009sq,Liu:2011re}. When 
applying in the study of the electron/positron excesses, such a global 
fitting method can fit both the background and the extra source parameters 
simultaneously and avoid the bias of choosing the background parameters. 
This approach definitely makes sense on the quantitative level, in spite 
that there are still uncertainties from various kinds of model configurations 
such as the CR propagation and the solar modulation \cite{Yuan:2014pka}.
It is expected that with better understandings of those issues based on
more and better data from AMS-02, the global fitting method may be
more powerful to probe the underlying physical nature of the CRs.

One potential problem of the previous studies about the CR leptons
is the systematical uncertainties among different detectors. As shown
by the preliminary data of AMS-02 presented in 2013 International
Cosmic Ray Conference \cite{2013ICRC-AMS02}, many kinds of measurements
showed differences compared with previous measurements. Furthermore,
the data-taking periods of various experiments are also different and
the solar modulation effect will be different. It is no longer a
problem after the most recent data release about the positron and
electron fluxes by AMS-02
\cite{AMS02-posi-2014,AMS02-elec-2014,AMS02-tot-2014}. In this work we
adopt the AMS-02 data about the positron fraction, positron plus
electron flux, positron flux and the electron flux to study the injection
properties of the backgrounds and the extra sources of the CR leptons.

This paper is organized as follows. We first give a description of our 
fitting process in Sec. II. The propagation of CRs in the Galaxy is 
introduced in \cref{section_the_propagation_of_cosmicrays_in_the_galaxy}.
The assumptions and parameterization of the backgrounds and extra sources
of electrons/positrons are described in 
\cref{section_the_background_$e^+$_and_$e^-$} and
\cref{section_the_extra_source}. The fitting results in different models
are given in \cref{section_fitting}. We give some discussions about the
results in \cref{section_discussion}, and conclude in \cref{section_conclusions}.

\section{The scheme of the global fitting}

The scheme of the global fitting follows our previous study of the AMS-02 
positron fraction results \cite{Yuan:2013eja}. The model is described by 
a set of parameters $\vec{\theta}$, which include the primary electron 
spectrum, the electron/positron spectrum from the extra sources such as 
pulsar-like astrophysical sources or the DM. These parameters will be 
defined in the next sections. Once the parameters are given we can calculate 
the propagation of the CRs in the Milky Way. The production and propagation 
of secondary positrons/electrons will also be calculated at the same time.
Then we compare the predicted spectra with the AMS-02 data and evaluate
the model by minimizing the $\chi^2$.

The MCMC technique is used to derive the posterior probability
distributions of the parameters from the observational data. According
to the Bayes theorem, the posterior probability of a set of parameters
$\vec{\theta}$ in light of the observational data is ${\cal P}(\vec{\theta}
|D)\propto{\cal P}(D|\vec{\theta}){\cal P}(\vec{\theta})$, where
${\cal P}(D|\vec{\theta})={\cal L}(\vec{\theta})\propto\exp(-\chi^2
(\vec{\theta})/2)$ is the likelihood function of model $\vec{\theta}$ 
for the data, and ${\cal P}(\vec{\theta})$ is the prior probability of the
model parameters before the current observations. In this work we 
adopt flat (constant) prior probabilities of all the model parameters
in specified ranges (some of them are logarithmical, see details in the 
tables below).

We adopt the Metropolis-Hastings algorithm to generate the Markov
chains from the unknown target distribution. The Metropolis-Hastings 
algorithm adopts a propose-and-accept process, in which the acceptance
or rejection of a proposed point depends on the probability ratio between
this point and the former one, to generate the chains. Such a sampling
method can still work efficiently when the dimension of the parameter 
space is high.

The propagation model parameters will be first determined by fitting
the B/C and $^{10}$Be/$^9$Be data (see more detailed description in
\cref{section_the_propagation_of_cosmicrays_in_the_galaxy}).
The propagation parameters are then fixed to be the best-fitting values
when fitting the lepton data. The proton injection spectrum is also
determined by fitting the AMS-02 data \cite{2013ICRC-AMS02}.
Therefore in the fitting process the parameter space $\vec{\theta}$
includes only the parameters of the lepton sector.

The global fitting gives us information of the background and properties
of the extra sources at the same time. Therefore results on the astrophysical
sources and DM are not biased due to the choice of background.
The parameters for DM given in the work can be taken as the starting point
for the future model-building. The results for pulsars can also be
a guideline for pulsar model study although the case for pulsars is more
complicated as each pulsar may have different properties.
What we get may indicate the property of a nearby pulsar which gives
dominant contribution to the positron excess.

\section{The propagation of cosmic rays in the Galaxy}
\label{section_the_propagation_of_cosmicrays_in_the_galaxy}
Galactic CR particles diffuse in the Galaxy after being accelerated,
suffering from the fragmentation and energy loss in the ISM and/or the
interstellar radiation field (ISRF) and magnetic field, decay and possible
reacceleration or convection. Denoting the density of CRs per unit
momentum interval as $\psi$, the propagation can be described by the
propagation equation
\begin{widetext}
  \begin{equation}
    \begin{split}
      \frac{\partial \psi}{\partial t} &= Q(\mathbf{x}, p) + \nabla \cdot \left( D_{xx}\nabla\psi - \mathbf{V}_{c}\psi \right)
      + \frac{\partial}{\partial p}p^2D_{pp}\frac{\partial}{\partial p}\frac{1}{p^2}\psi
      - \frac{\partial}{\partial p}\left[ \dot{p}\psi - \frac{p}{3}\left( \nabla\cdot\mathbf{V}_c\psi \right) \right]
      - \frac{\psi}{\tau_f} - \frac{\psi}{\tau_r},
    \end{split}
    \label{propagation_equation}
  \end{equation}
\end{widetext}
where $Q(\mathbf{x}, p)$ is the source distribution, $D_{xx}$ is the spatial
diffusion coefficient, $\mathbf{V}_c$ is the convection velocity, $D_{pp}$
is diffusion coefficient in the momentum-space, $\tau_f$ and $\tau_r$ are
the characteristic time scales used to describe the fragmentation and
radioactive decay. The convection velocity $\mathbf{V}_c$ is generally
assumed to linearly depend on the distance away from the Galaxy disk.
The diffusion coefficient can be parameterized as $D_{xx} = D_0\beta
\left( R/R_0 \right)^{\delta}$, where $\beta$ is the velocity of the
particle in unit of light speed $c$, $R\equiv pc/Ze$ is the rigidity.
The reacceleration effect is described with the diffusion in momentum
space. Considering the scenario in which the CR particles are reaccelerated
by colliding with the interstellar random weak hydrodynamic waves, the
relation between the spatial diffusion coefficient $D_{xx}$ and the
momentum diffusion coefficient $D_{pp}$ can be expressed as
\cite{1990acr..book.....B,1994ApJ...431..705S}:
\begin{equation}
  D_{pp}D_{xx}=\frac{4p^2v^2_{A}}{3\delta(4-\delta^2)(4-\delta)\omega},
  \label{reacceleration}
\end{equation}
where $v_{A}$ is the Alfven velocity and the parameter $\omega$ is used
to characterize the level of the interstellar turbulence. Since only
$v^2_A/\omega$ is relevant, we adopt $\omega = 1$ and refer $v_A$ to
characterize the reacceleration. Free escape is assumed at the boundaries,
$R_h$ and $z_h$, for cylindrical geometry.

The secondary-to-primary ratios of nuclei are almost independent of the
injection spectrum. They are always employed to constrain the propagation
parameters in Eq. (\ref{propagation_equation}). Generally used are the
Boron-to-Carbon ratio (B/C) and unstable-to-stable Beryllium ratio
($^{10}$Be/$^9$Be). The B/C ratio is sensitive to the average path of the
CR particles go through between the source and the observer, which
correlate positively with both $D_{xx}$ and the diffusion halo size $z_h$.
The $^{10}$Be/$^9$Be ratio is sensitive to probe the resident time of
particles in the Galaxy, which correlate positively with $z_h$ but negatively
with $D_{xx}$. Therefore, combining these two ratios, the main propagation
properties can be fixed.

The major parameters to describe the propagation are $(D_0, \delta,
v_A, \mathrm{d}V/\mathrm{d}z, z_h)$. Since there are degeneracies between
the models with reacceleration and convection effects, and the current
data of B/C and $^{10}$Be/$^9$Be are not effective enough to distinguish
them, we adopt two distinct scenarios as benchmark models of the propagation.
They are referred as diffusion reacceleration (DR) model and diffusion
convection (DC) model, respectively.

\begin{figure*}[!htbp]
  \centering
  \includegraphics[width=0.46\textwidth]{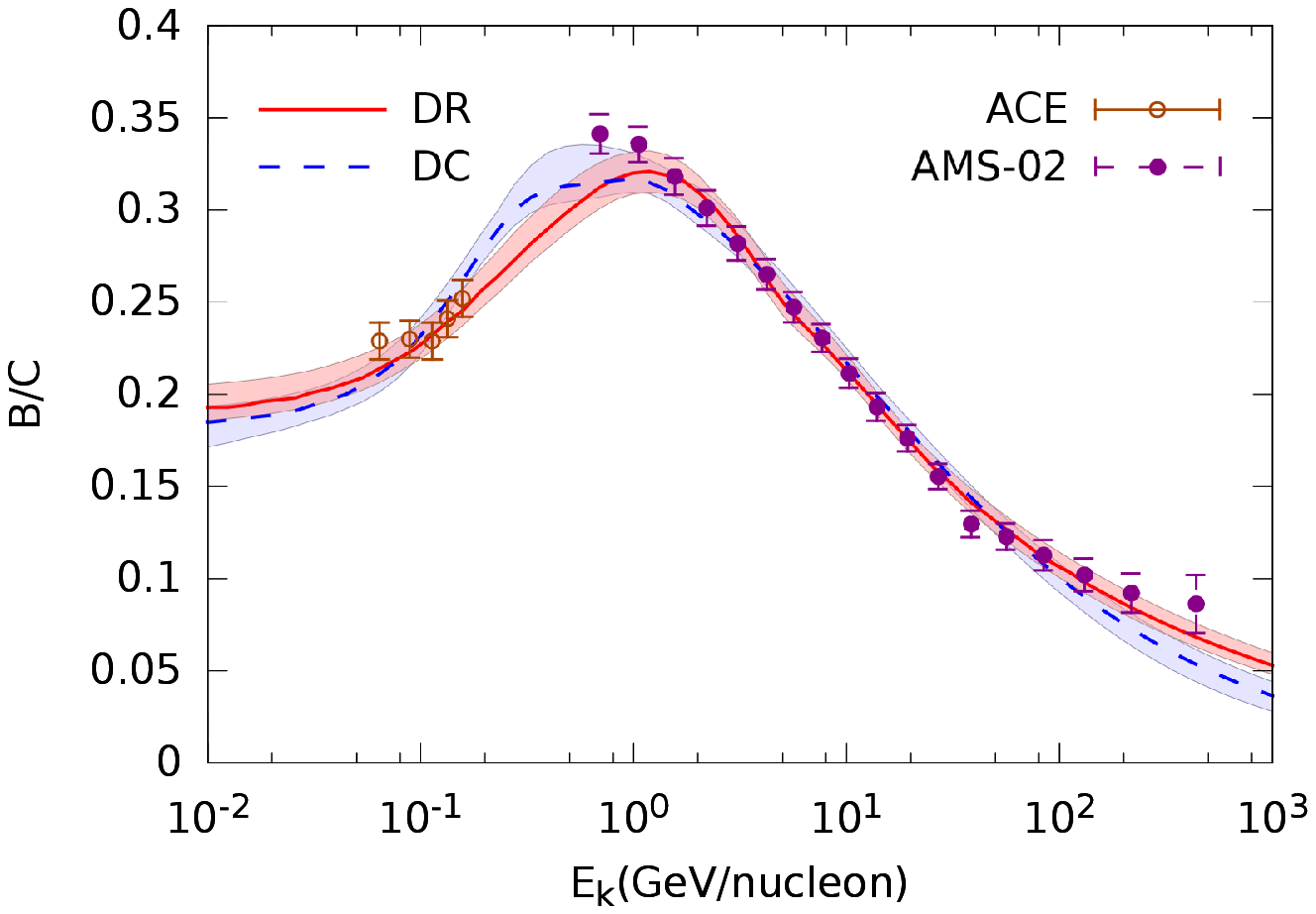}
  \includegraphics[width=0.46\textwidth]{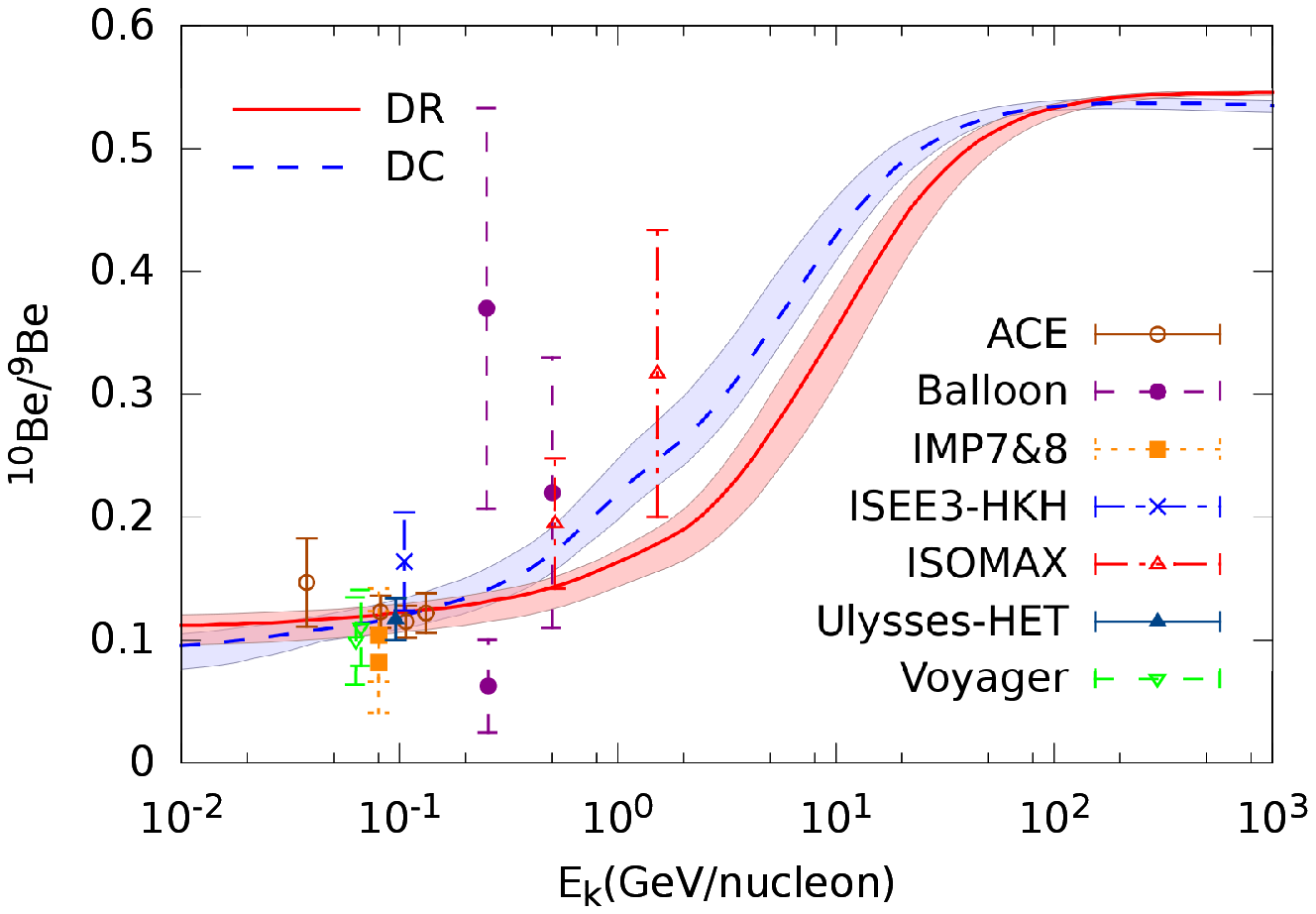}
  \caption{The B/C ratio (left) and $^{10}\mathrm{Be}/^{9}\mathrm{Be}$
  ratio (right) for the corresponding parameters shown in Table
  \ref{tab:propagation_coefficient}, compared with the data. The bands
  show the $95\%$ confidence ranges. The B/C data are from AMS-02 
  \cite{2013ICRC-AMS02} and ACE 
  \cite{2000AIPC..528.....M}, and the $^{10}$Be/$^9$Be data are from
  experiments ACE \cite{2001ApJ...563..768Y},
  Balloon \cite{1977ApJ...212..262H,1978ApJ...226..355B,1979ICRC....1..389W}, 
  IMP7\&8 \cite{1981ICRC....2...72G}, ISEE3-HKH \cite{1988SSRv...46..205S}, ISOMAX
  \cite{2004ApJ...611..892H}, Ulysses-HET \cite{1998ApJ...501L..59C} and
  Voyager \cite{1999ICRC....3...41L}.}
  \label{fig:BC_Be}
\end{figure*}

The public numerical tool, GALPROP version 54.1.984\footnote{Available at
http://galprop.stanford.edu/} \cite{Strong:1998pw,Moskalenko:1997gh}, is
adopted to calculate the propagation of CR particles. We employ the B/C
data from AMS-02 \cite{2013ICRC-AMS02} and ACE \cite{2000AIPC..528.....M},
and the $^{10}$Be/$^9$Be data from experiments ACE \cite{2001ApJ...563..768Y},
Balloon \cite{1977ApJ...212..262H,1978ApJ...226..355B,1979ICRC....1..389W}, 
IMP7\&8 \cite{1981ICRC....2...72G}, ISEE3-HKH \cite{1988SSRv...46..205S},
ISOMAX \cite{2004ApJ...611..892H}, Ulysses-HET \cite{1998ApJ...501L..59C}
and Voyager \cite{1999ICRC....3...41L} to constrain the propagation
parameters. The MCMC method is adopted to fit the B/C and $^{10}$Be/$^9$Be
data. In order to reproduce the low energy B/C data, a broken power law,
where $\delta$ is 0 when $R$ is below $R_0$, is adopted for $D_{xx}$ in
the DC scenario \cite{Moskalenko:2001ya}. To describe the propagation of
CR particles in the solar system, we adopt the force-field approximation
\cite{Gleeson:1968zza}, which contains only one free parameter, the
so-called solar modulation potential $\phi$. The mean values and $1\sigma$
errors of the propagation parameters for DR and DC scenarios are shown 
in Table \ref{tab:propagation_coefficient}. Fig. \ref{fig:BC_Be}
shows the fitting results of B/C (left) and $^{10}$Be/$^9$Be (right) ratios
within $2\sigma$ confidence level, compared with the observational data.

\begin{table}
  \centering
  \caption{The mean values and $1\sigma$ uncertainties of the propagation 
parameters derived through fitting the B/C and $^{10}$Be/$^9$Be ratios. 
In the DC scenario $\delta$ is set to be 0 when $R$ is below $R_0$.}
  \begin{tabular}{*{4}{>{$}l<{$}}}
    \hline
    \hline
                            &                                   &  \mathrm{DR}    &  \mathrm{DC}\\
    \hline
    D_0                     &  (10^{28}\cm^2\s^{-1})     &  6.58\pm1.27    &  1.95\pm0.50\\
    \delta                  &                                   &  0.333\pm0.011  &  0.510\pm0.034\\
    R_0                     &  (\GV)                    &  4              &  4.71\pm0.8\\
    v_A                     & (\km\,\s^{-1})         &  37.8\pm2.7     &  \text{---}\\
    \mathrm{d}V/\mathrm{d}z & (\km\,\s^{-1}\kpc^{-1}) &  \text{---}     &  4.2\pm3.2\\
    z_h                     & (\kpc)                    &  4.7\pm1.0      &  2.5\pm0.7\\
    \phi                    & (\MV)                     &  326\pm36       &  182\pm25\\
    \hline
    \hline
  \end{tabular}
  \label{tab:propagation_coefficient}
\end{table}

\section{The parameters for the background $e^+$ and $e^-$ spectra}
\label{section_the_background_$e^+$_and_$e^-$}

Electrons are also expected to be accelerated during the acceleration of
CR nuclei at the sources, e.g. supernova remnants (SNRs). During the
propagation, the inelastic collision between the nuclei and the ISM will
produce secondary electrons and positrons. These components consist of
the background contribution of electrons and positrons. Such a picture
is supported by the observations of secondary-to-primary ratio of nuclei
as well as the diffuse $\gamma$-ray emission \cite{Strong:2007nh,
2012ApJ...750....3A}.

The spatial distribution of the injected CR particles is assumed to
follow the SNR distribution
\begin{equation}
f(r,z) = \left( \frac{r}{r_\odot} \right)^a \exp\left( -b\cdot\frac{r-
r_\odot}{r_\odot} \right)\exp\left( -\frac{\left|z\right|}{z_s} \right),
\label{spatial_distribution}
\end{equation}
where $r_\odot=8.5$ kpc is the distance from the Sun to the Galactic
center, $z_s\approx0.2$ kpc is the characteristic height of Galactic disk.
The two parameters $a$ and $b$ are chosen to be 1.25 and 3.56 following
\cite{Trotta:2010mx}, which are adjusted to fit the $\gamma$-ray gradient.
The injection spectra of all kinds of nuclei are assumed to be a broken
power law form
\begin{equation}
  q_\mathrm{i}=  N_\mathrm{i}\times\left\{ \begin{array}{ll}
    \left( \dfrac{R}{R\mathrm{_{br}^p}} \right)^{-\nu_1} & R \le R\mathrm{_{br}^p}\\
    \left( \dfrac{R}{R\mathrm{_{br}^p}} \right)^{-\nu_2} & R > R\mathrm{_{br}^p}
  \end{array}
  \right.,
  \label{injection_power_law}
\end{equation}
where $\mathrm{i}$ denote species of the nuclei, $R$ is the rigidity of
the particle, and $N_\mathrm{i}$ is the normalization constant proportional
to the relative abundance of the corresponding nuclei. The injection is
simply assumed to be stable, which is an effective approximation if the
production rate of SNRs is high enough. Therefore, the injection source
function can then be written as $Q_\mathrm{i}({\bf x},p)=f(r,z)
q_\mathrm{i}(p)$.

\begin{figure}[!htp]
  \centering
  \includegraphics[width=0.47\textwidth]{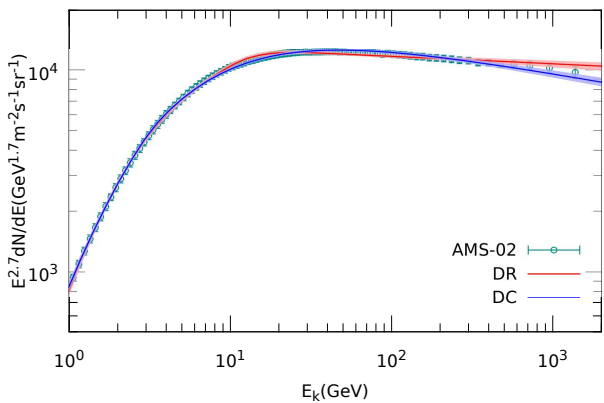}
  \caption{The fluxes of protons for the corresponding parameter shown in
Table \ref{tab:proton_injection}, compared with the preliminary data from
AMS-02 \cite{2013ICRC-AMS02}. The bands delimit the regions of 95\% 
confidence level.}
  \label{fig:proton_flux}
\end{figure}

Adopting the propagation parameters as the best-fitting values shown in
Table \ref{tab:propagation_coefficient}, we then constrain the injection
parameters of Eq. (\ref{injection_power_law}) with the proton flux of
AMS-02 \cite{2013ICRC-AMS02}. The resulting nuclei injection parameters
are given in Table \ref{tab:proton_injection}. Fig. \ref{fig:proton_flux}
shows the best-fitting results and the 95\% confidence ranges of the 
proton fluxes, compared with the AMS-02 measurement.

The secondary production of electrons and positrons can then be calculated
with the propagated proton (and Helium) spectra. We use the parameterization
presented in \cite{Kamae:2006bf} to calculate the production spectrum of
secondary electrons and positrons. To partially take into account the
uncertainties when calculating the secondary fluxes, from e.g.,
proton-proton collision cross section, enhancement factor from heavier
nuclei, and/or the propagation uncertainties, we employ a parameter
$c_{e^+}$ to re-scale the calculated secondary flux to fit the data. Note
that the above mentioned uncertainties may not be simply represented
with a constant factor, but most probably they are energy dependent
\cite{Delahaye:2008ua,Mori:2009te}. Here a constant factor is just an
approximation and the purpose is to fit the data.

\begin{table}
  \centering
  \caption{The nucleon injection parameters derived through fitting
the proton data of AMS-02. A single power-law is enough to fit the data
in the DC scenario.}
\begin{tabular}{>{$}l<{$}cpp}
    \hline
    \hline
                             & Prior Range & \text{DR} & \text{DC} \\
    \hline
    \nu_1                    &   [1.0, 4.0]   & 1.811, 0.021 & 2.336, 0.004 \\
    \nu_2                    &   [1.0, 4.0]   & 2.402, 0.005 & 2.336, 0.004 \\
    R_\mathrm{br}^p(\GV)     &   [8.0, 15.0]  & 12.88, 0.263 & 10.00 \\
    A_p\footnote{Post-propagated normalization flux of protons at 100 GeV in unit $10^{-9}\cm^{-2}\s^{-1}\sr^{-1}\MeV^{-1}$}
                             &   [3.0, 6.0]   & 4.613, 0.027 & 4.783, 0.026 \\
    \phi_p(\MV)              &   [50, 1500]   & 517.8, 37.8  & 505.9, 13.1 \\
    \hline
    \hline
  \end{tabular}
  \label{tab:proton_injection}
\end{table}

About the primary electrons, we also assume a broken power-law form of the
injection spectrum. Since electrons lose energies much more efficiently than
the nuclei, the effect from recent and nearby sources may make the spectrum
fluctuate significantly, especially at high energies \cite{DiMauro:2014iia}.
Therefore there might be more structures in the electron spectrum.
We will discuss two cases in this work: one break case similar to Eq.
(\ref{injection_power_law}), and three-piece broken power law with two
breaks. The latter is found to be required to fit the pre-AMS-02 lepton
data \cite{Feng:2013zca,Yuan:2013eba,Cholis:2013psa,Yuan:2014pka}.
Thus the electron injection parameters are
\begin{equation}
  \begin{array}{ll}
    \text{with one break:} & (\gamma_1, \gamma_2, R^e_\mathrm{br}, A_e);\\
    \text{with two breaks:} & (\gamma_1, R^e_\mathrm{br}, \gamma_2, R^e_\mathrm{br2}, \gamma_3, A_e).\\
  \end{array}
\end{equation}

\section{The parameters for the extra sources}
\label{section_the_extra_source}

In this paper, two kinds of extra source, including pulsars and the DM
annihilation, will be discussed. The pulsars are able to generate high
energy positron-electron pairs through the electromagnetic cascade in the
magnetic pole region, which could cause the observed excess
\cite{1970ApJ...162L.181S,2001A&A...368.1063Z,Yuksel:2008rf,Hooper:2008kg,
Profumo:2008ms}. The injection spectrum of the electrons and positrons
is usually assumed to be a power law with an exponential cutoff
\begin{equation}
  q_e^{\mathrm{psr}} = A_\mathrm{psr}(R/\mathrm{MV})^{-\alpha}\mathrm{exp}
  \left(-R/R_\mathrm{c} \right),
  \label{pulsar_injection}
\end{equation}
where $A_{\mathrm{psr}}$ is the normalization factor, $\alpha$ is the
spectral index, $R_\mathrm{c}$ is the cutoff rigidity. We adopt a continuous
and stable pulsar injection. The spatial distribution obeys the same form
of Eq. (\ref{spatial_distribution}), with slightly different parameters
$a=2.35$ and $b=5.56$ \cite{2004IAUS..218..105L}.

The particle and antiparticle of DM in the Galaxy, if the interaction
is strong enough, can also annihilate with each other and produce
standard model particles which are injected in the Galaxy as CRs.
Since there is no obvious excess of antiprotons from the secondary
expectation during the CR propagation compared with the data
\cite{Adriani:2010rc}, leptonical annihilation final states are expected
\cite{Cirelli:2008pk,Yin:2008bs}. We therefore discuss the model with
annihilation final states of a pair of muons or tauons. 
We use the results of PPPC 4 DM ID \cite{Cirelli:2010xx}, which
includes the electroweak corrections \cite{Ciafaloni:2010ti}, to calculate 
the electron (positron) spectrum from DM annihilation.
The Navarro-Frenk-White (NFW) density profile \cite{Navarro:1996gj}
\begin{equation}
  \rho(r)=\frac{\rho_s}{\left( r/r_s \right)\left( 1+r/r_s \right)^2},
  \label{NFW_profile}
\end{equation}
where $\rho_s = 20\kpc$ and $\rho_s=0.26\GeV\cm^{-3}$, is adopted to
describe the spatial distribution of DM in the Milky Way halo.
The free parameters in the DM annihilation scenario include the DM
particle mass $m_\chi$ and the velocity-weighted average annihilation
cross section $\langle \sigma v\rangle$.

\section{The fitting results} \label{section_fitting}

The datasets used in this study include the latest measurements of 
the positron fraction $e^+/(e^++e^-)$, fluxes of $e^-$, $e^+$ and 
($e^++e^-$) by AMS-02 \cite{AMS02-posi-2014,AMS02-elec-2014,AMS02-tot-2014}.
These data may not be fully uncorrelated, since the positron fraction may 
be derived from the fluxes of $e^-$ and $e^+$. However, the analysis methods 
are different for various kinds of measurements, and the systematic 
uncertainties also differ from one to another. Therefore we adopted all 
these data in the study, though the statistics of the data would be over 
estimated. We have also tested that when dropping one group of the data 
the results are almost unchanged. We further select the data above
1 GeV, since the lowest energy data may be significantly affected by the
solar modulation and may not be well modeled in the force-field
approximation \cite{Gleeson:1968zza}. The spectral index of
electrons/positrons injected by pulsars, and possibly altered by the
surrounding pulsar wind nebulae, is actually very uncertain. In this work
we limit $\alpha$ to the range between 1.0 and 2.4, according to the radio
and $\gamma$-ray observations of pulsars \cite{1988ApJ...327..853R,
Thompson:1994sg,Fierro:1995ny}.

\subsection{One break in primary electron spectrum}
\label{subsection_with_one_break_in_primary_injection}

\begin{figure*}[!htp]
  \centering
  \includegraphics[width=0.48\textwidth]{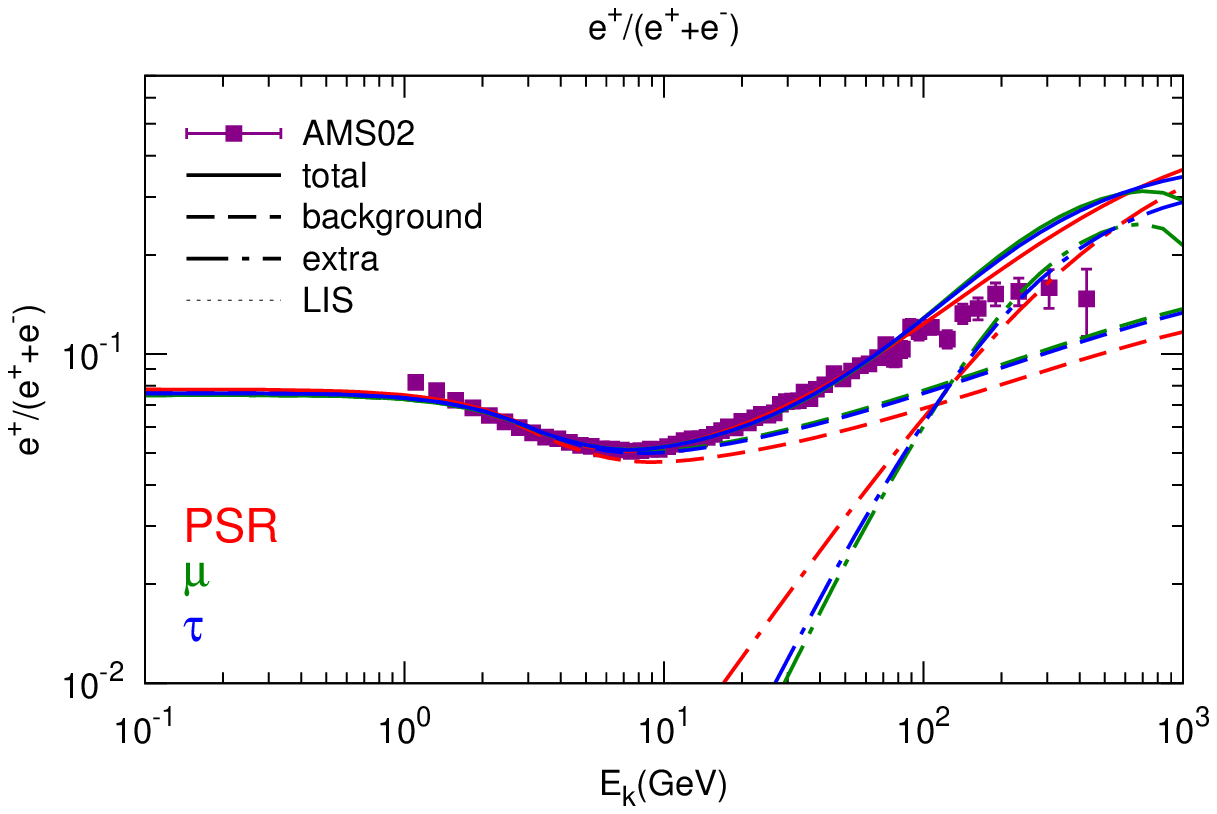}
  \includegraphics[width=0.48\textwidth]{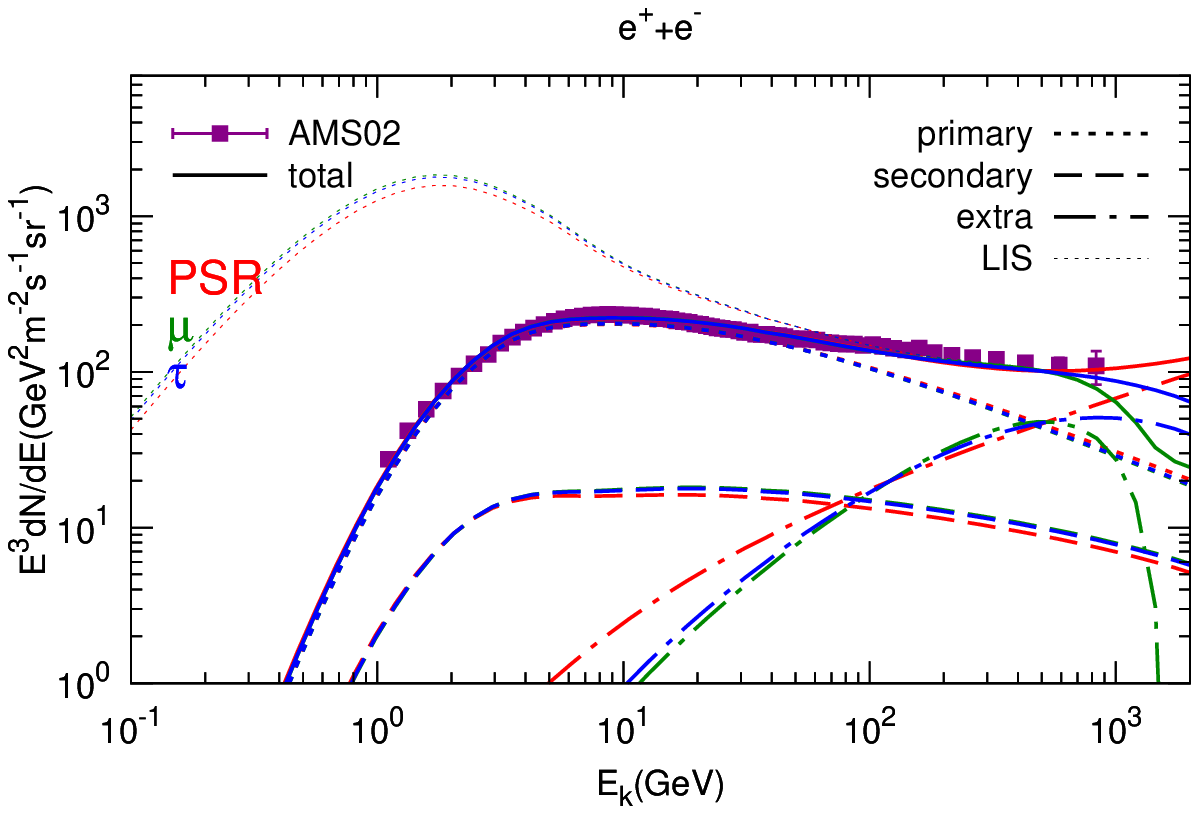}\\
  \includegraphics[width=0.48\textwidth]{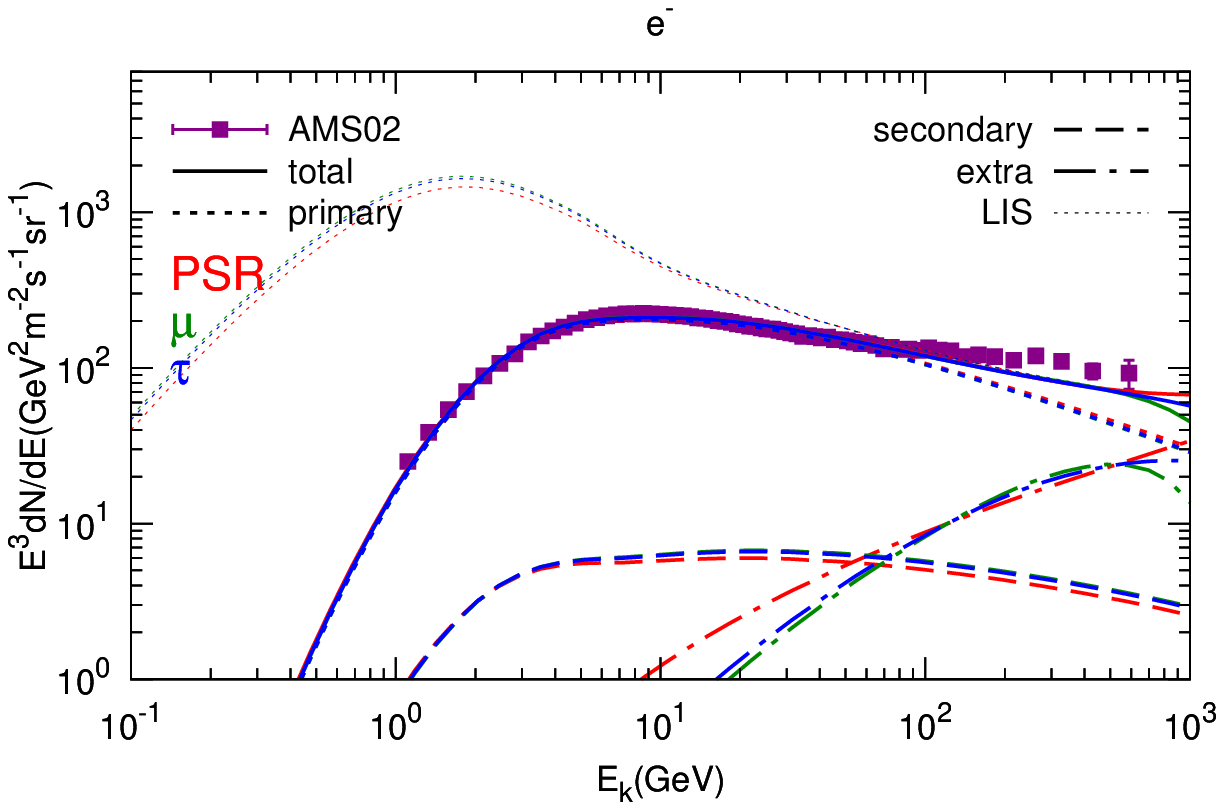}
  \includegraphics[width=0.48\textwidth]{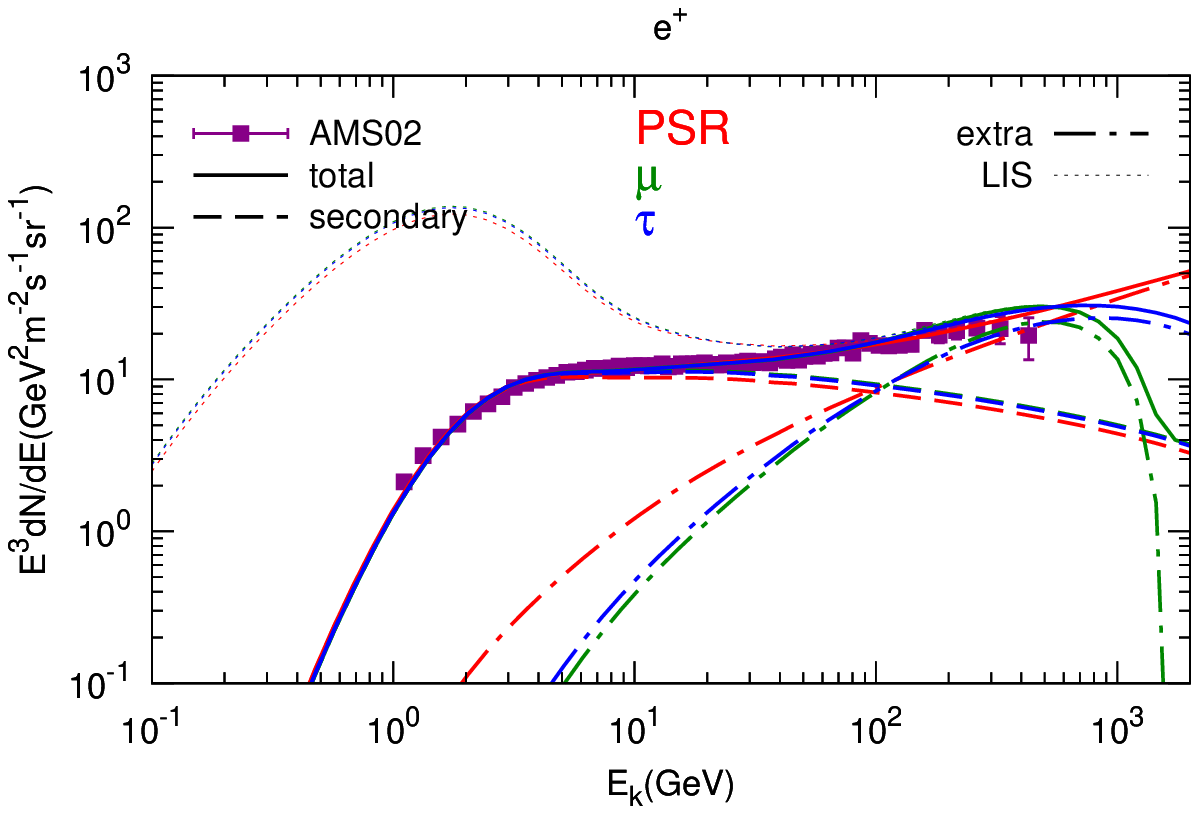}
  \caption{The expected results for the best-fitting parameters in DR
scenario with one-break of the primary electron injection spectrum.
Top-left: positron fraction $e^+/(e^++e^-)$; top-right: electron plus
positron flux $e^++e^-$; bottom-left: electron flux $e^-$; bottom-right:
positron flux $e^+$. The red, green and blue curves represent the pulsar,
DM annihilation into $\mu^+\mu^-$ and $\tau^+\tau^-$ final states,
respectively. Different line styles represent different components as
labeled.}
  \label{fig:DRnbk}
\end{figure*}

\begin{figure*}[!htp]
  \centering
  \includegraphics[width=0.48\textwidth]{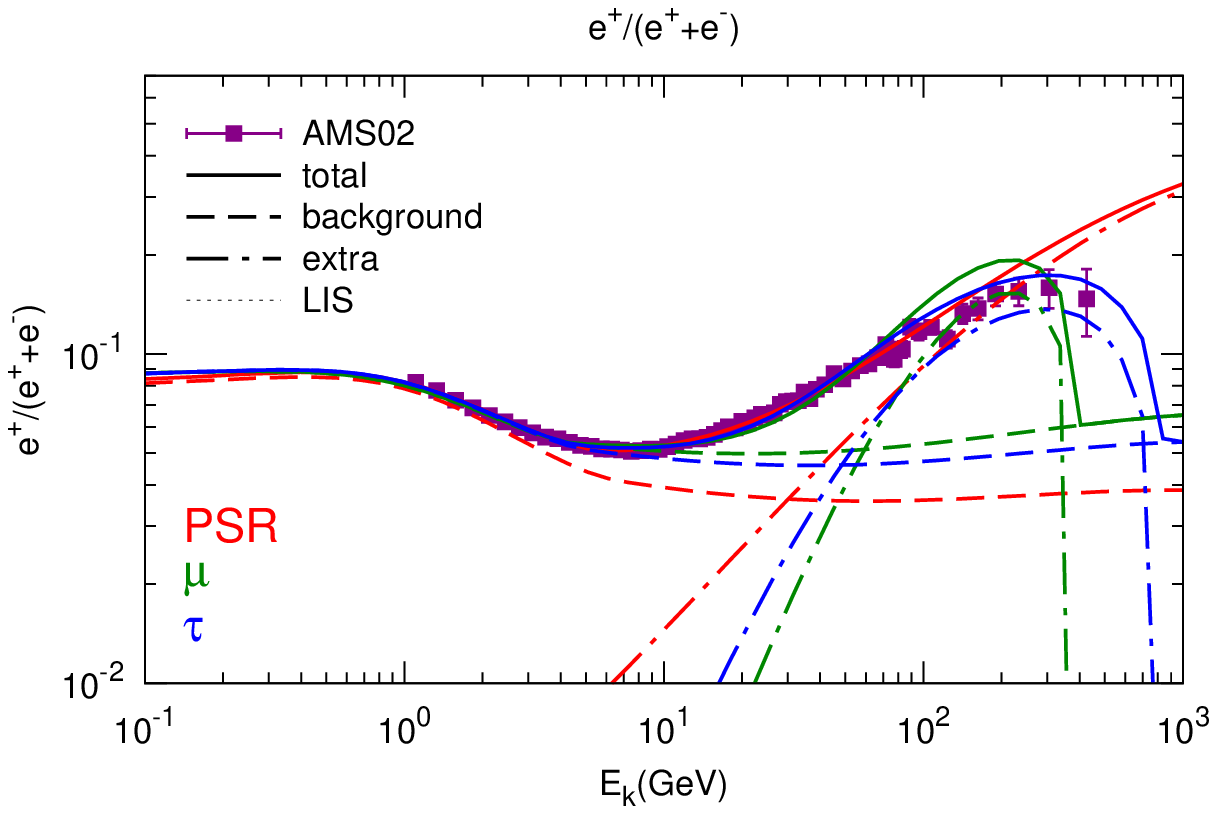}
  \includegraphics[width=0.48\textwidth]{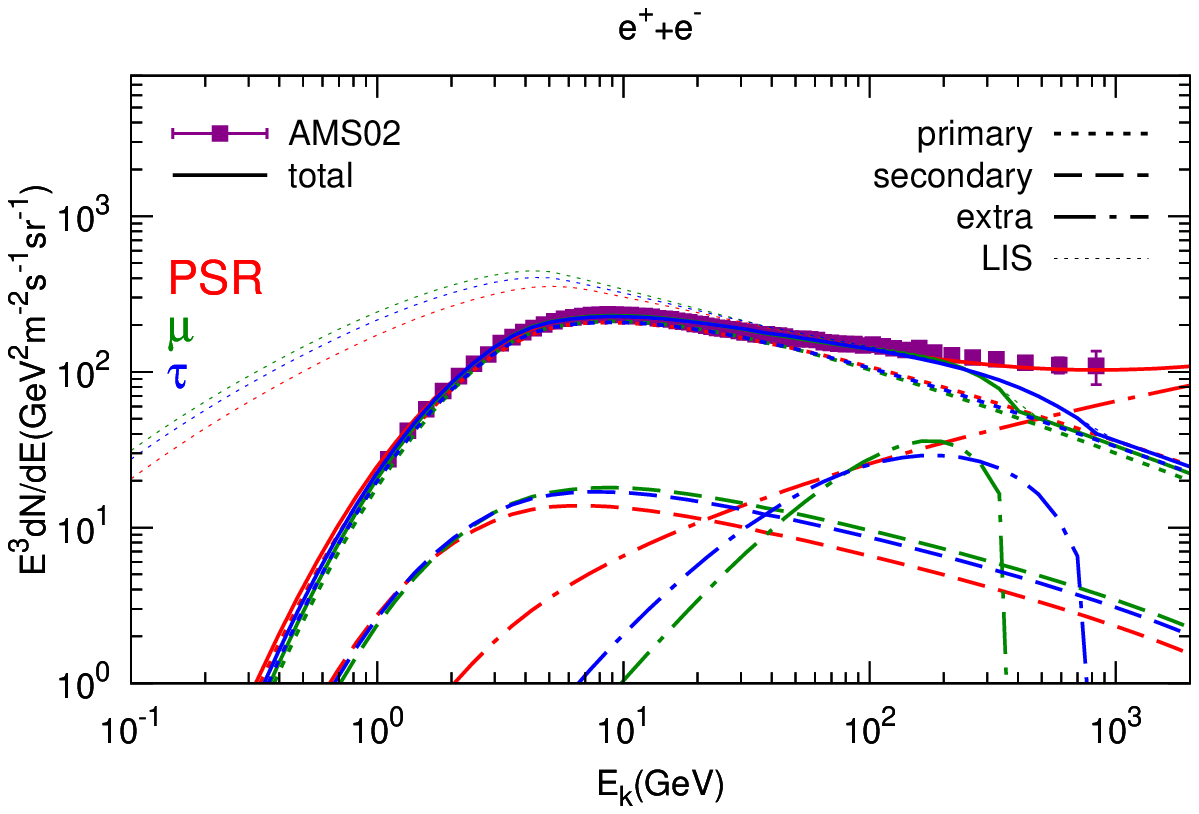}\\
  \includegraphics[width=0.48\textwidth]{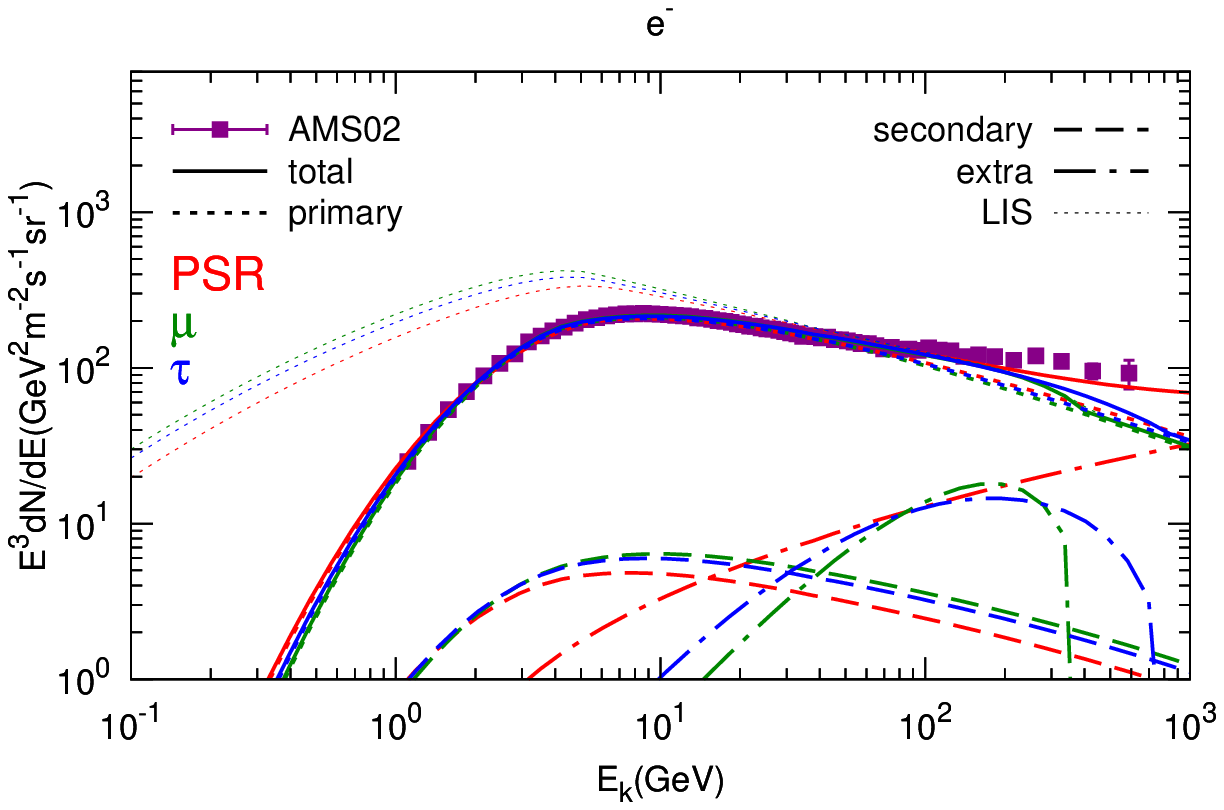}
  \includegraphics[width=0.48\textwidth]{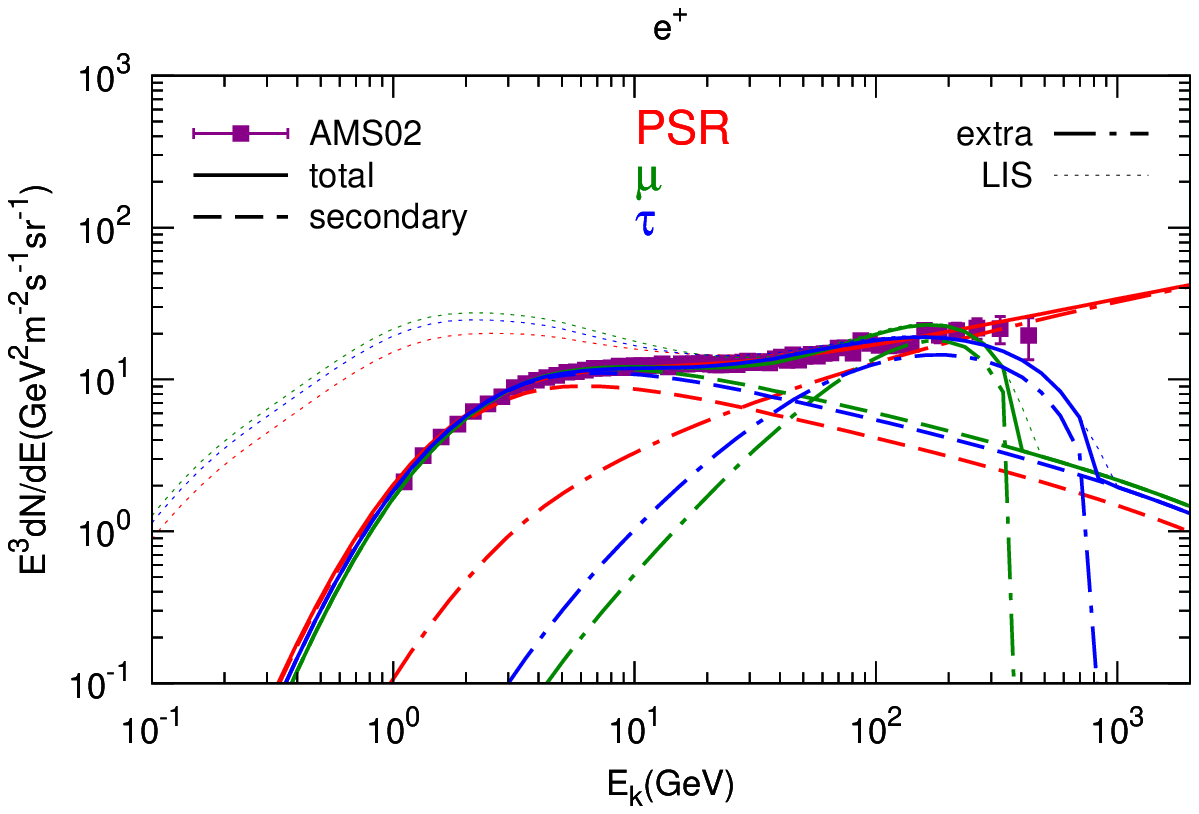}
  \caption{Same as Fig. \ref{fig:DRnbk} but for DC propagation scenario.}
  \label{fig:DCnbk}
\end{figure*}

In this case there are 4 parameters of the primary electrons. Together with
$c_{e^+}$, $\phi$ and the extra source parameters, we have in total 9 (8)
parameters for the pulsar (DM) scenario. The best-fitting parameters and
the mean values as well as $1\sigma$ confidence ranges are shown in
Tables \ref{tab:PSRnbk}, \ref{tab:MUnbk} and \ref{tab:TAUnbk} for the
pulsar, DM annihilation into $\mu^+\mu^-$ and DM annihilation into
$\tau^+\tau^-$ respectively. Figs. \ref{fig:DRnbk} and \ref{fig:DCnbk}
illustrate the comparison of the best-fitting results with the data,
for DR and DC propagation scenarios. Table \ref{tab:chi2} summarizes the
fitting $\chi^2$ values for each dataset.

\begin{table*}
  \centering
  \caption{Fitting results of pulsar model with one break in $e^-$ injection spectrum}
  \include{tab/paraPSRnbk}
  \label{tab:PSRnbk}
\end{table*}

\begin{table*}
  \centering
  \caption{Fitting results of DM annihilation scenario in $\mu^+\mu^-$ channel with one break in $e^-$ injection spectrum}
  \include{tab/paraMUnbk}
  \label{tab:MUnbk}
\end{table*}

\begin{table*}
  \centering
  \caption{Fitting results of DM annihilation scenario in $\tau^+\tau^-$ channel with one break in $e^-$ injection spectrum}
  \include{tab/paraTAUnbk}
  \label{tab:TAUnbk}
\end{table*}

\begin{table*}
  \centering
  \caption{Fitting $\chi^2$ values and the contribution from each data.
The number of data points for $e^+/(e^++e^-)$, $e^++e^-$, $e^-$ and $e^+$ are
63, 71, 70 and 69, respectively.}
  \include{tab/chi2tab}
  \label{tab:chi2}
\end{table*}

The results show not good enough fittings to data. From Figs. \ref{fig:DRnbk}
and \ref{fig:DCnbk} it can be seen that while the model may over-produce
the positron fraction, it is not enough to reproduce the electron flux at
high energies. This is similar to that we found before using the electron
data from PAMELA/Fermi-LAT \cite{Yuan:2013eja}. The fitting $\chi^2$ values
also show this issue. The minimum $\chi^2$ value for these 6 fittings
is $374.2$, and the reduced $\chi^2$ is about $1.42$ for $264$ degree of
freedom (d.o.f.). It corresponds to a $4.4\sigma$ deviation from a good
fitting as expected.

Given the fittings are poor, the constraints on the model parameters
by minimizing the $\chi^2$ may not be physically meaningful. The pulsar
model gives better fitting than the DM scenario, since the spectral index
of electrons/positrons injected by pulsar is enabled to vary and it has larger
d.o.f. compared with the spectrum expected from DM annihilation. We further
note that the $\chi^2$ contributed from the $e^-$ flux and the
$e^++e^-$ flux are about two times larger than that from $e^+$ flux,
although the numbers of data points are comparable. It could be due to the
fact that electrons have much higher statistics compared with the
positrons. The failure to reproduce the high energy electron spectrum well
would result in a large $\chi^2$ value.  Therefore, we may need to change
the background model to improve the fitting of the high energy electron
spectrum.

\subsection{Two breaks in primary electron spectrum}
\label{subsection_with_two_break_in_primary_injection}

\begin{figure*}[!htp]
  \centering
  \includegraphics[width=0.48\textwidth]{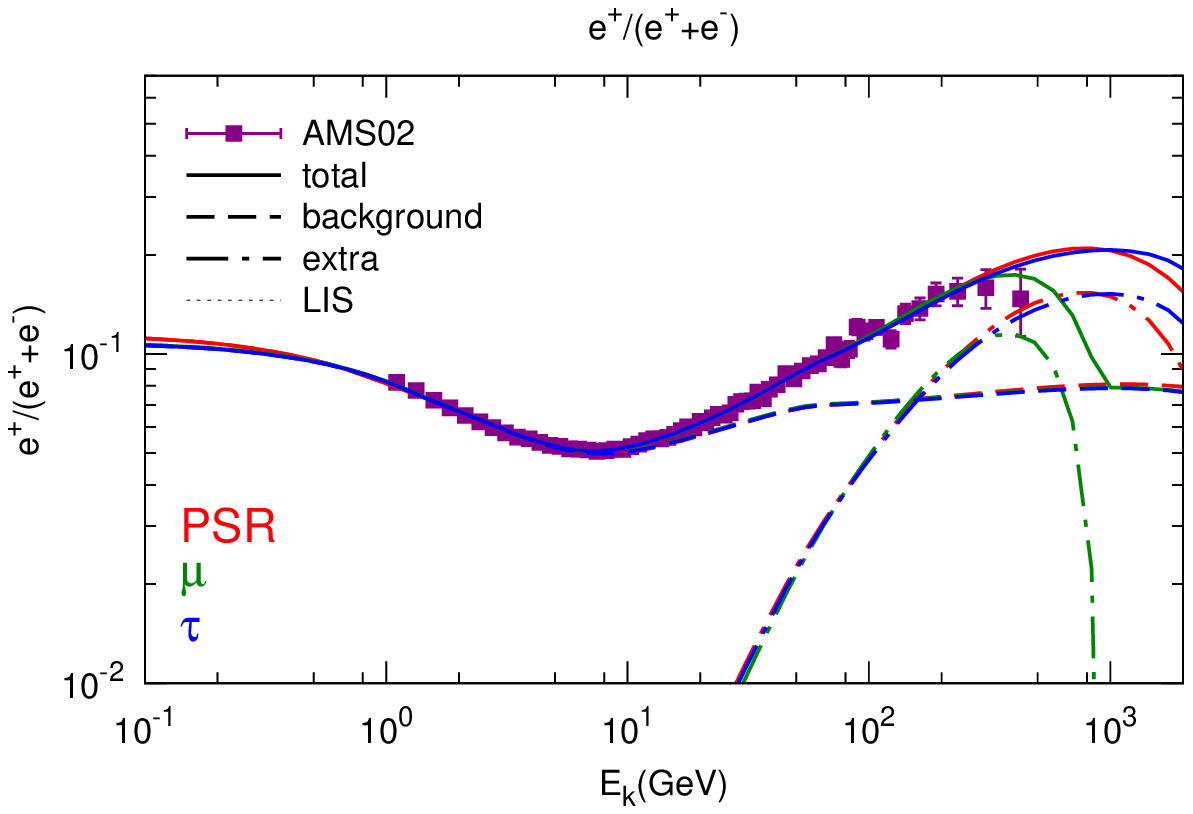}
  \includegraphics[width=0.48\textwidth]{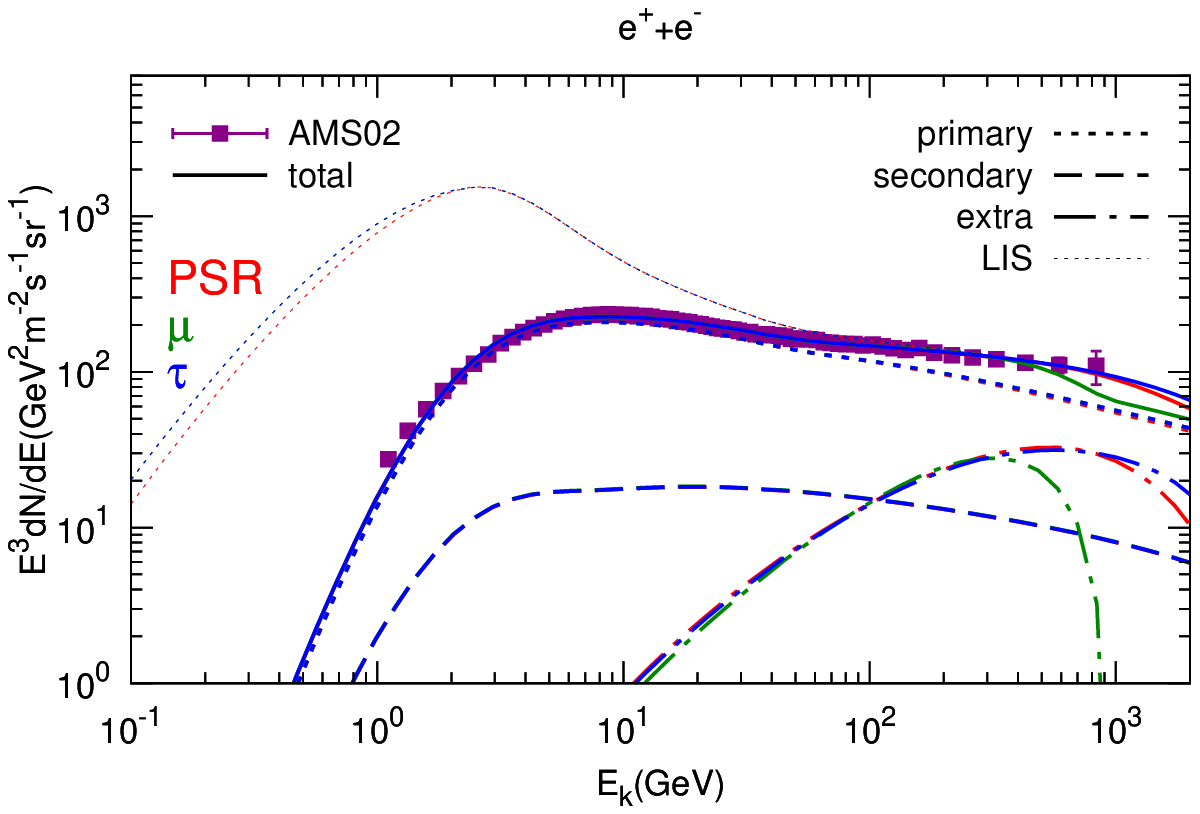}\\
  \includegraphics[width=0.48\textwidth]{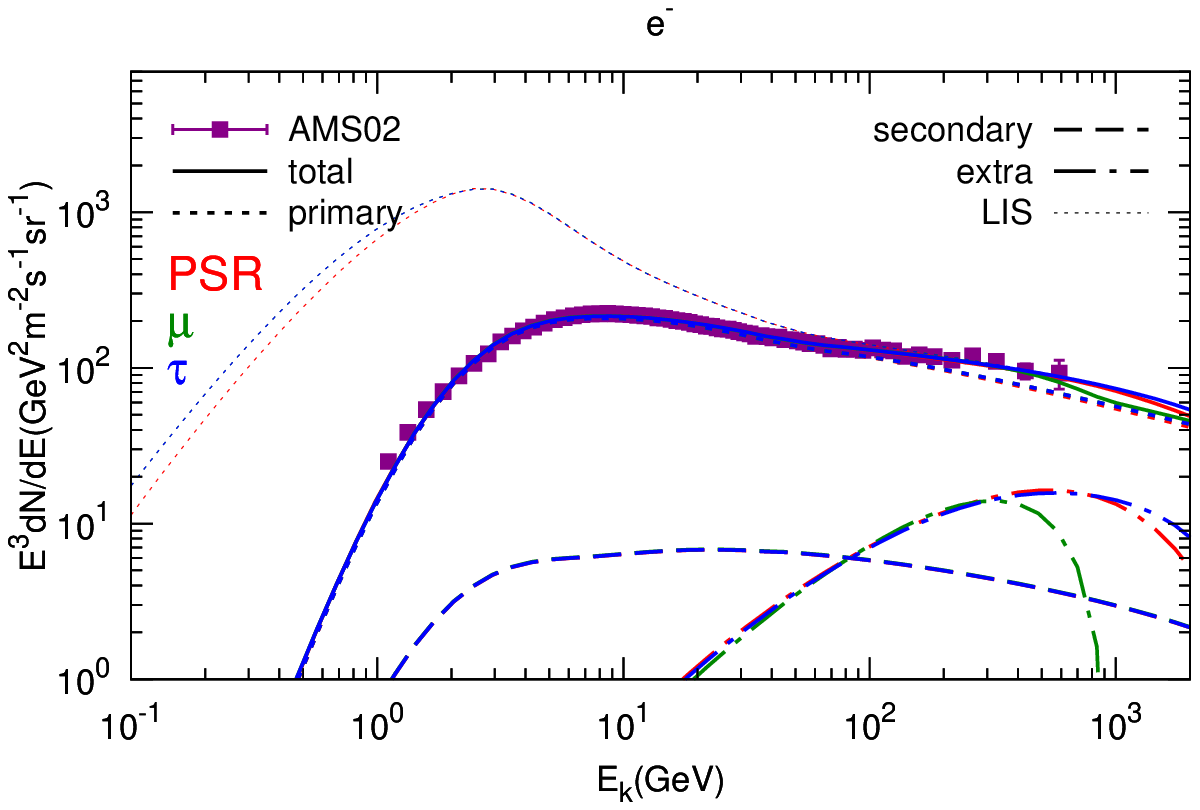}
  \includegraphics[width=0.48\textwidth]{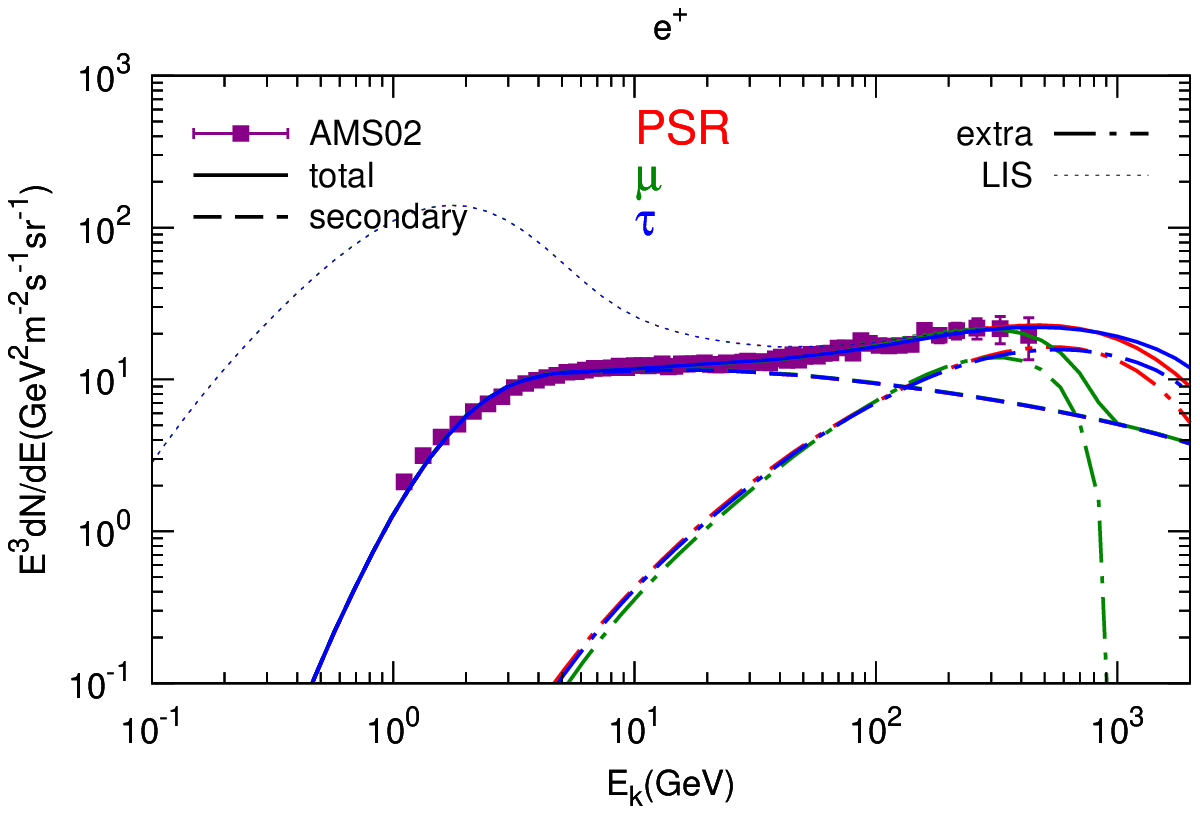}
  \caption{The same as Fig. \ref{fig:DRnbk} but for the background model
with two breaks in the electron injection spectrum.}
  \label{fig:DRbk}
\end{figure*}

\begin{figure*}[!htp]
  \centering
  \includegraphics[width=0.48\textwidth]{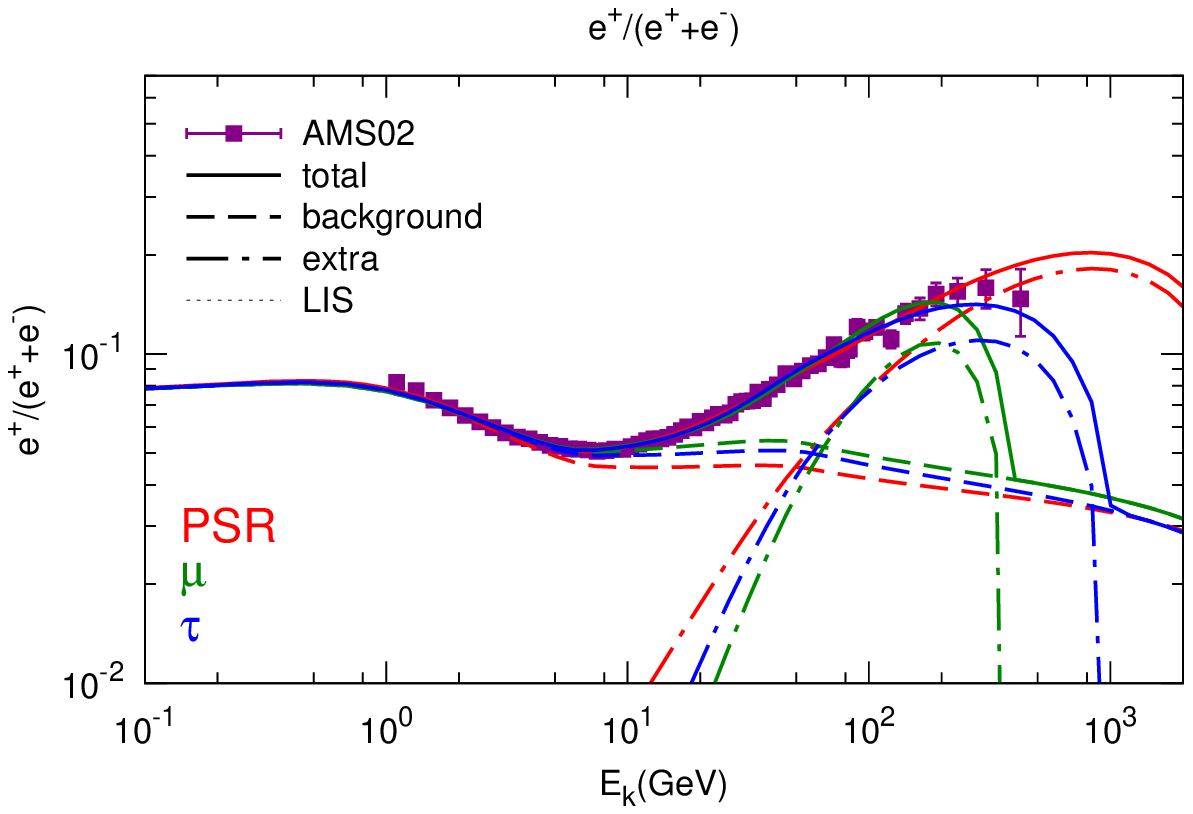}
  \includegraphics[width=0.48\textwidth]{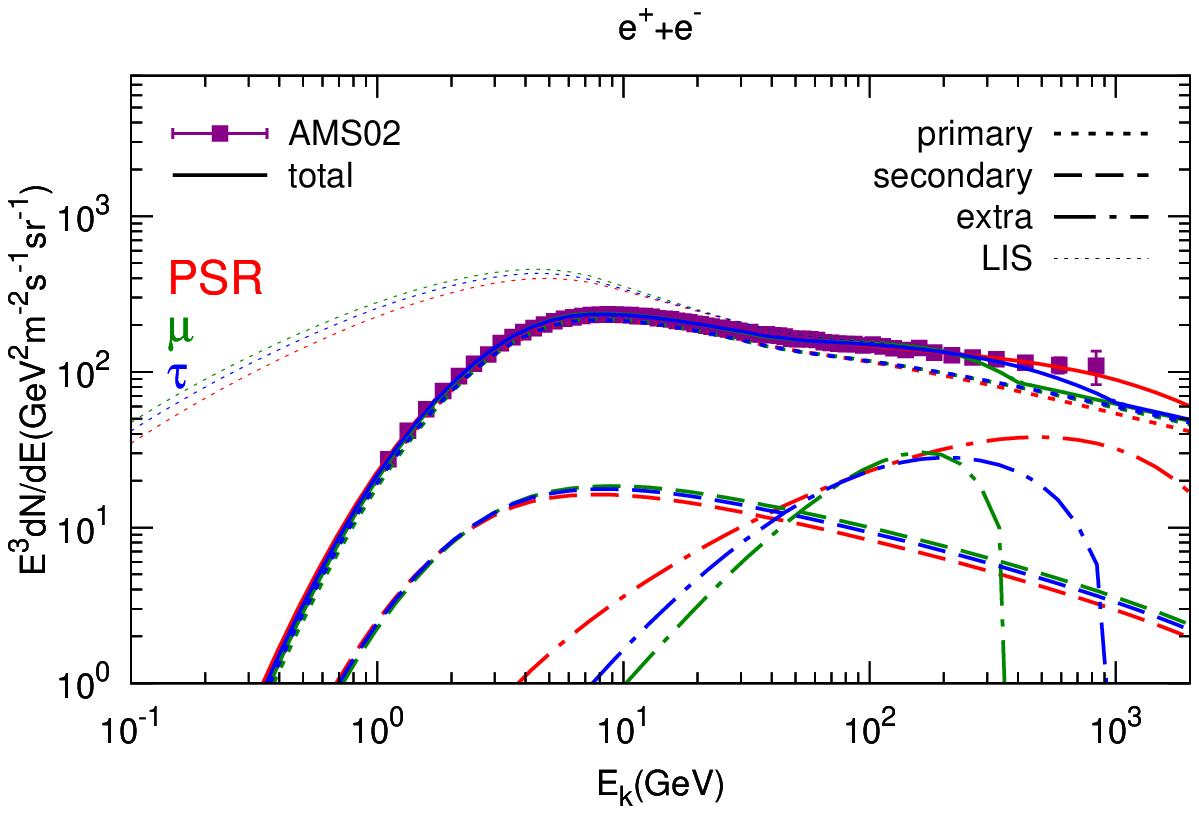}\\
  \includegraphics[width=0.48\textwidth]{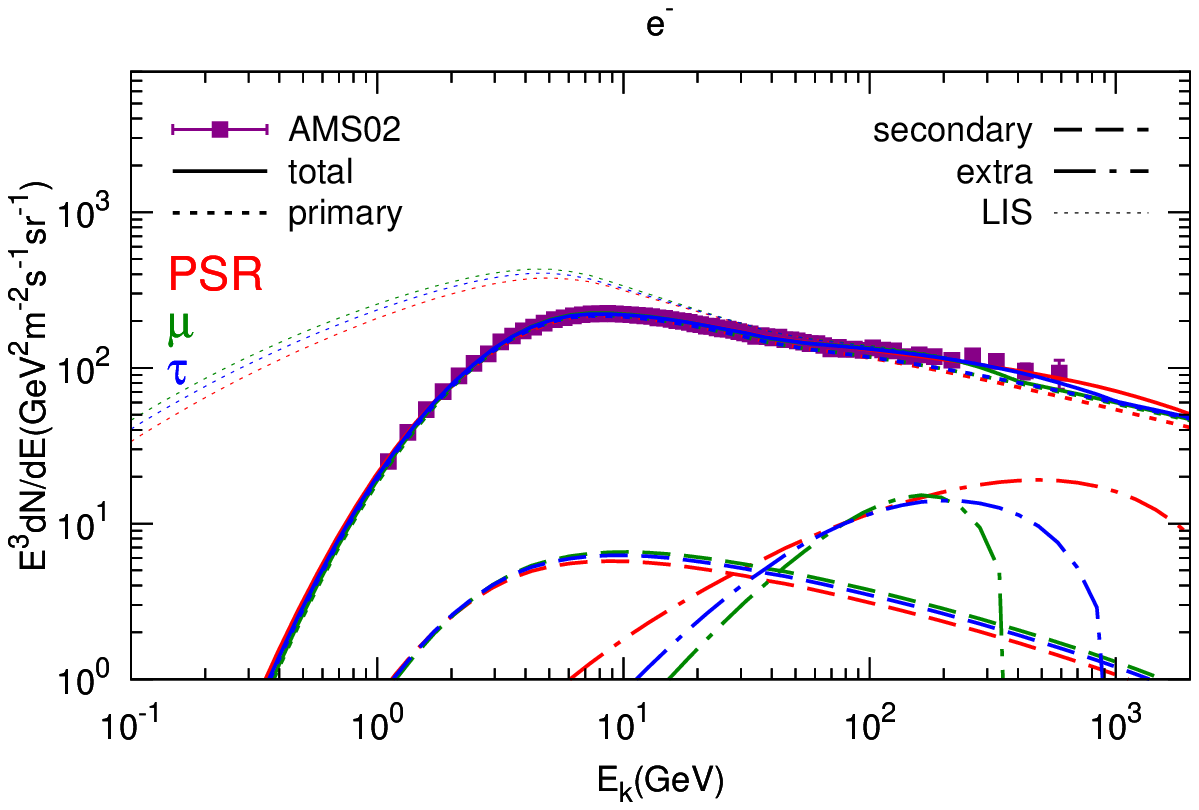}
  \includegraphics[width=0.48\textwidth]{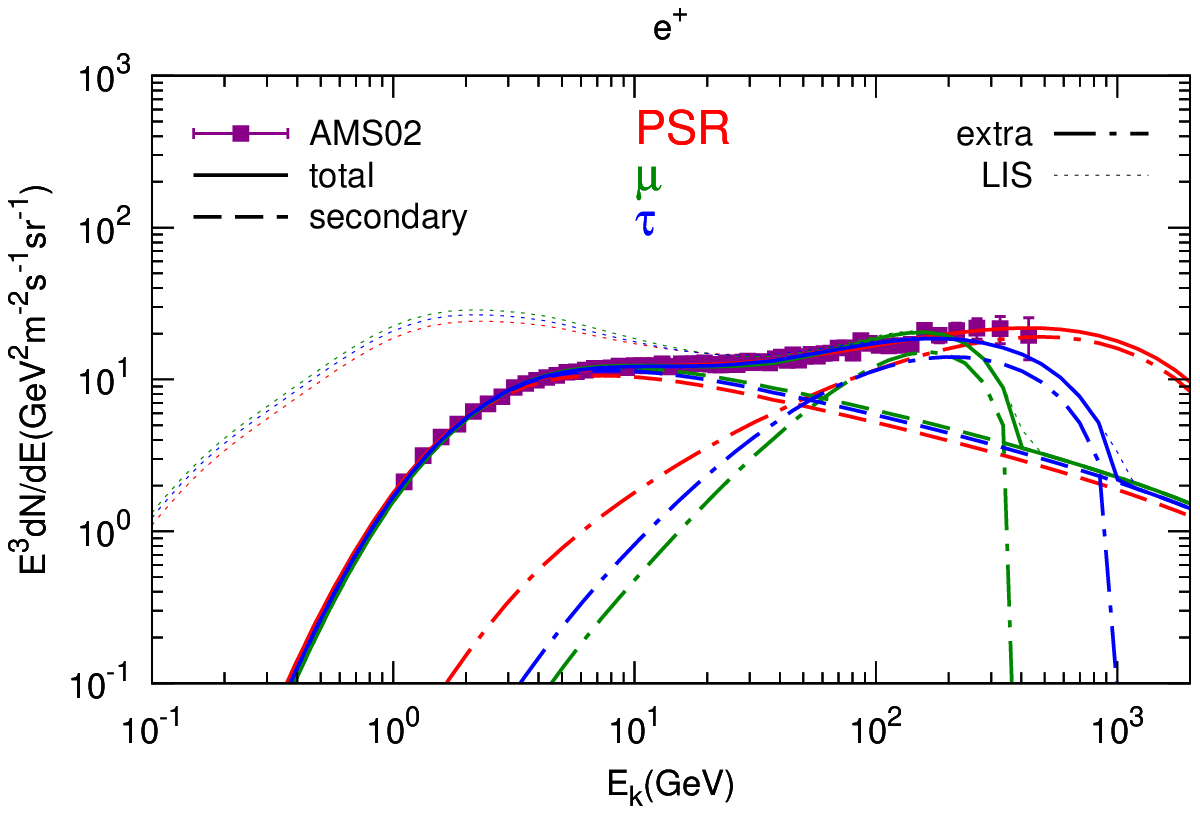}
  \caption{The same as Fig. \ref{fig:DRbk} but for DC propagation scenario.}
  \label{fig:DCbk}
\end{figure*}

\begin{figure*}[!htp]
  \centering
  \includegraphics{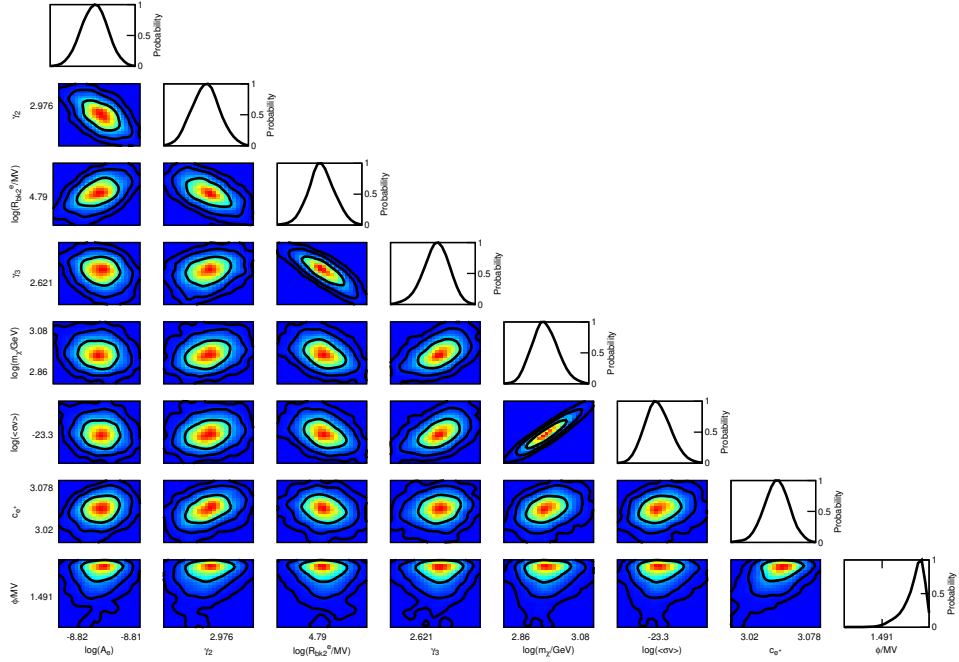}
  \caption{The 1 and 2-dimensional distributions of part of the parameters 
  for DM annihilation into $\mu^+\mu^-$ channel. The background electrons 
  have two breaks, and the propagation model is DR. The contours in the 
  2-dimensional plots denote the 68.3\%, 95.5\% and 99.7\% confidence levels 
  from inside to outside.
}
  \label{fig:triangle_plot_DR}
\end{figure*}

\begin{figure*}[!htp]
  \centering
  \includegraphics{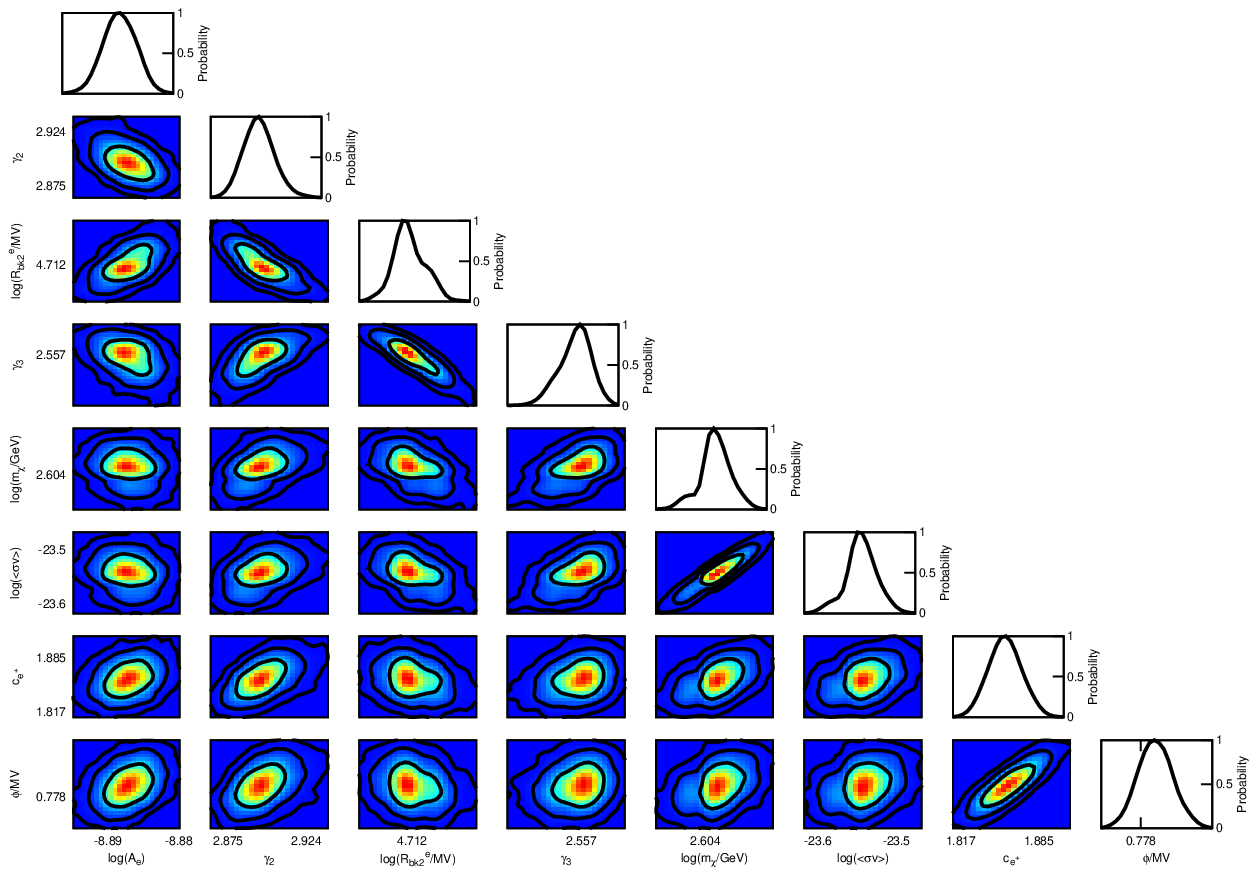}
  \caption{The same as Fig. \ref{fig:triangle_plot_DR}, but for the DC
  propagation scenario.
}
  \label{fig:triangle_plot_DC}
\end{figure*}

A direct way to alleviate the tension shown above is to add more
electrons at high energies, such as a spectral hardening \cite{Feng:2013zca,
Cholis:2013psa,Yuan:2013eba}. For the nuclei spectra similar spectral
hardening above several hundred GV has been observed by ATIC
\cite{Panov:2006kf}, CREAM \cite{Ahn:2010gv} and PAMELA \cite{Adriani:2011cu},
and could be naturally expected if there is a diversity of the source
parameters \cite{Yuan:2011ys}. Therefore we apply a second break on the
injection spectrum of the primary electrons characterized by two additional
parameters $\gamma_3$ and $R_{\rm br2}^e$. The fitting results are shown
in Tables \ref{tab:PSRbk}, \ref{tab:MUbk}, \ref{tab:TAUbk} and Figs.
\ref{fig:DRbk}, \ref{fig:DCbk} respectively. We also show in  
Figs. \ref{fig:triangle_plot_DR} and \ref{fig:triangle_plot_DC} the
derived one and two dimentional posterior distributions of the most 
relevant parameters, choosing the DM annihilation into $\mu$ channel as 
benchmark cases.

\begin{table*}
  \centering
  \caption{Fitting results of pulsar model with two breaks in $e^-$ injection spectrum}
  \include{tab/paraPSRbk}
  \label{tab:PSRbk}
\end{table*}

\begin{table*}
  \centering
  \caption{Fitting results of DM annihilation scenario in $\mu^+\mu^-$ channel with two breaks in $e^-$ injection spectrum}
  \include{tab/paraMUbk}
  \label{tab:MUbk}
\end{table*}

\begin{table*}
  \centering
  \caption{Fitting results of DM annihilation scenario in $\tau^+\tau^-$ channel with two breaks in $e^-$ injection spectrum}
  \include{tab/paraTAUbk}
  \label{tab:TAUbk}
\end{table*}

Significant improvements of the fittings can be seen from these results.
It is shown from Table \ref{tab:chi2} that in all the cases the reduced
$\chi^2$ values are about 2 times smaller than the previous case with
one break. For most cases the reduced $\chi^2$ is close to or
smaller than 1.
Since the systematic errors are added quadratically to the statistical
errors to calculate the $\chi^2$, it is expected that the reduced $\chi^2$
value will be smaller than 1 if the model does fit the data well (see e.g.
the minimum model of \cite{AMS02-elec-2014}). It can be seen from Figs.
\ref{fig:DRbk} and \ref{fig:DCbk} that the contribution of positrons from
the extra source dominates over the secondary component above $\sim 50-100$
GeV. For electrons, however, the background component will always dominate
in the energy range from 1 GeV to 1 TeV.

\begin{figure}[!htp]
  \centering
  \includegraphics[width=0.48\textwidth]{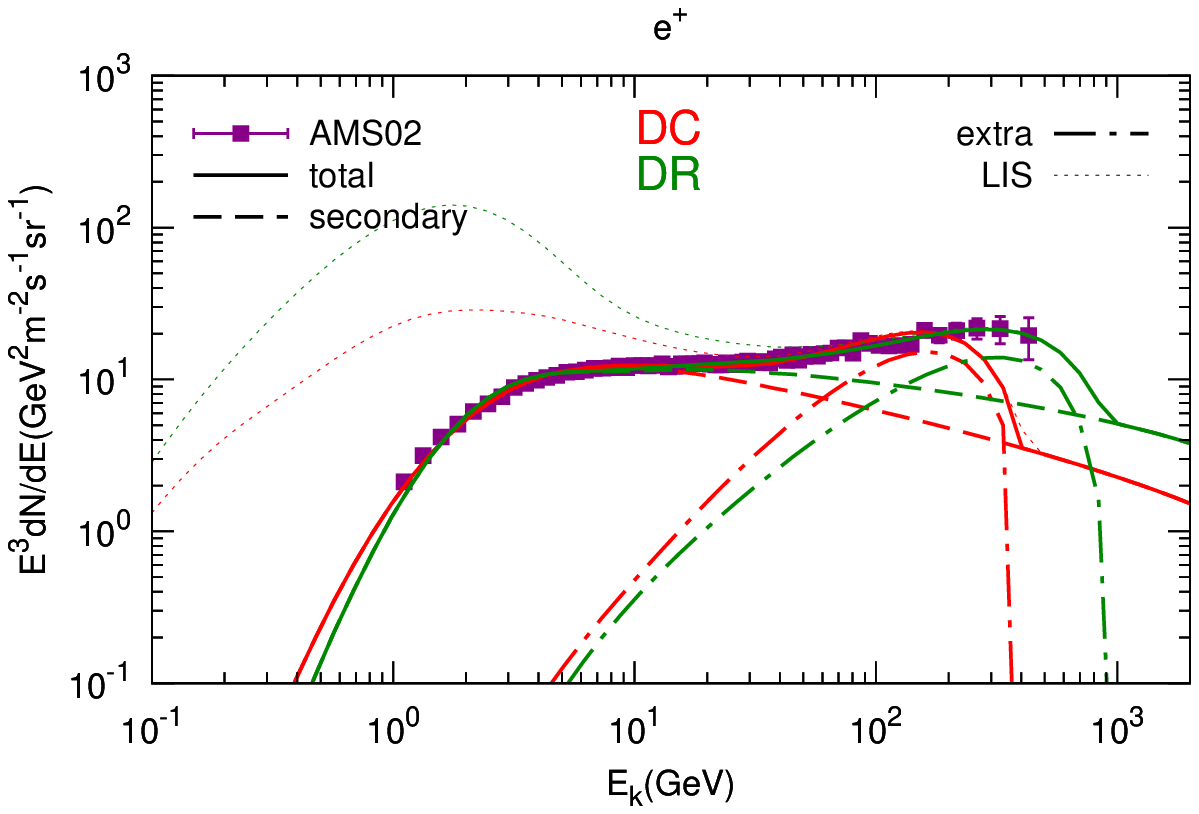}
  \caption{Comparison of the positron spectra between the DC scenario (red) 
and DR scenario (green) with extra source component from DM annihilation 
into $\mu^+\mu^-$ final state.}
  \label{fig:compare_DC_DR}
\end{figure}

We further find that the pulsar model and DM annihilation into 
$\tau^+\tau^-$ give comparable fittings to the data. However, the case 
for $\mu^+\mu^-$ channel seems to be more complicated. In the DR scenario 
it gives comparable fittings compared with the pulsar model and the DM 
annihilation into $\tau^+\tau^-$ channel, while in the DC scenario, the 
fitting results become worse. Fig. \ref{fig:compare_DC_DR} shows the 
comparison of the positron fluxes between DR and DC scenarios, for DM 
annihilation into $\mu^+\mu^-$ channel. The reason of such a result might 
be the difference of the secondary positron spectrum in the two scenarios.
As can be seen from Fig. \ref{fig:compare_DC_DR} the secondary positron 
spectrum is softer in the DC scenario. This should be due to that
the DC model has a larger propagation parameter $\delta$ than the
DR model. Since the positron spectrum from muon decay is very hard, 
a smaller value of DM mass is needed to better fit the data. It results 
in the failure to reproduce the high energy tail of the positron spectrum. 
The positron spectrum from pulsar or tauon decay can be softer, thus can 
fit the data better than the muon model\footnote{A four-muon final state
may give better fitting to the AMS-02 data.}.

\section{Discussion} \label{section_discussion}

In this paper we try to give a quantitative study on the AMS-02 results
of the electron/positron fluxes. Although the AMS-02 data are precise
enough, there are large uncertainties from the theoretical model parameters,
such as the uncertainties from the CR propagation model, the treatment of
the solar modulation, the Galactic gas distribution and so on.
In \cite{Yuan:2014pka}, we studied quite a few such kinds of uncertainties
as possible systematical uncertainties, including the propagation, the solar
modulation and low energy data selection, the hadronic interaction model
and so on. The study shows that although the uncertainties of the model
inputs seem to be large, the fitting results about the extra sources are
under good control.

As an illustration, we plot the $1\sigma$ and $2\sigma$ contours on the
$m_\chi-\langle \sigma v\rangle$ parameter plane to show the uncertainties
of the parameter determination for the scenarios discussed in this work.
The solid (dashed) ones are for the case with two (one) breaks of the
primary electron injection spectrum. The red ones are for DR propagation
model and blues ones are for DC model. The results do show some differences,
between DR and DC propagation models. Nevertheless, the shift of the
central values as well as the contours is about a factor of 4, which
is larger than that found in \cite{Yuan:2014pka}. One possible reason for
this difference might be that we do not include the HESS data at higher
energies in this study. The HESS data, although have large systematic
uncertainties, should be useful to constrain the very high energy behavior
of the electron/positron spectra. The future experiments such as 
DAMPE\footnote{http://dpnc.unige.ch/dampe/} and HERD \cite{Zhang:2014qga} 
may provide better measurements of the electron/positron spectra above TeV.

A main result of our fitting is that a new feature at the primary electron 
spectrum is strongly favored. Such a feature indicates that the nearby 
and/or fresh CR sources may contribute to the high energy electrons with 
a harder spectrum than the background \cite{DiMauro:2014iia}. 
Considering the large fluctuation of the electron/positron fluxes in 
space due to fast energy losses it is quite reasonable that the high 
energy electrons are dominated by the local sources. One possible signature 
of such a scenario may be the fine structures of the electron/positron 
spectra which may be investigated with future observations \cite{Yin:2013vaa}. 
Another possible probe of the local sources could be the anisotropy
measurements of the electrons \cite{Linden:2013mqa}.

Another interesting conclusion of this study is that in general 
the DC model is more favored than the DR model by the lepton spectra 
(Table \ref{tab:chi2}). The DM model with $\mu^+\mu^-$ channel is an 
exception which we have discussed above. The reason is that the local 
interstellar (LIS) spectrum has a bump at low energy in the
DR model due to the reacceleration, and thus a large solar modulation
potential has to be introduced to suppress the bump to fit the data.
Such a bump is necessary to better fit the B/C data 
\cite{Moskalenko:2001ya}, especially the HEAO data
\cite{1990A&A...233...96E}. However, the AMS-02 data about B/C do not
strongly favor a bump and hence the reacceleration for $E_k\lesssim1$
GeV/nucleon \cite{2013ICRC-AMS02}. The current data about B/C by AMS-02
is not able to distinguish DR from DC model. Therefore to finally
address the question that whether the reacceleration is favored
one needs more precise measurement of the B/C ratio down to sub-GeV.
Similar conclusion has also been obtained in the study of the synchrotron
radiation \cite{Strong:2011wd}. It was found that the DR model predicted
higher radio emission than observed.

\begin{figure*}[!htp]
  \centering
  \includegraphics[width=0.48\textwidth]{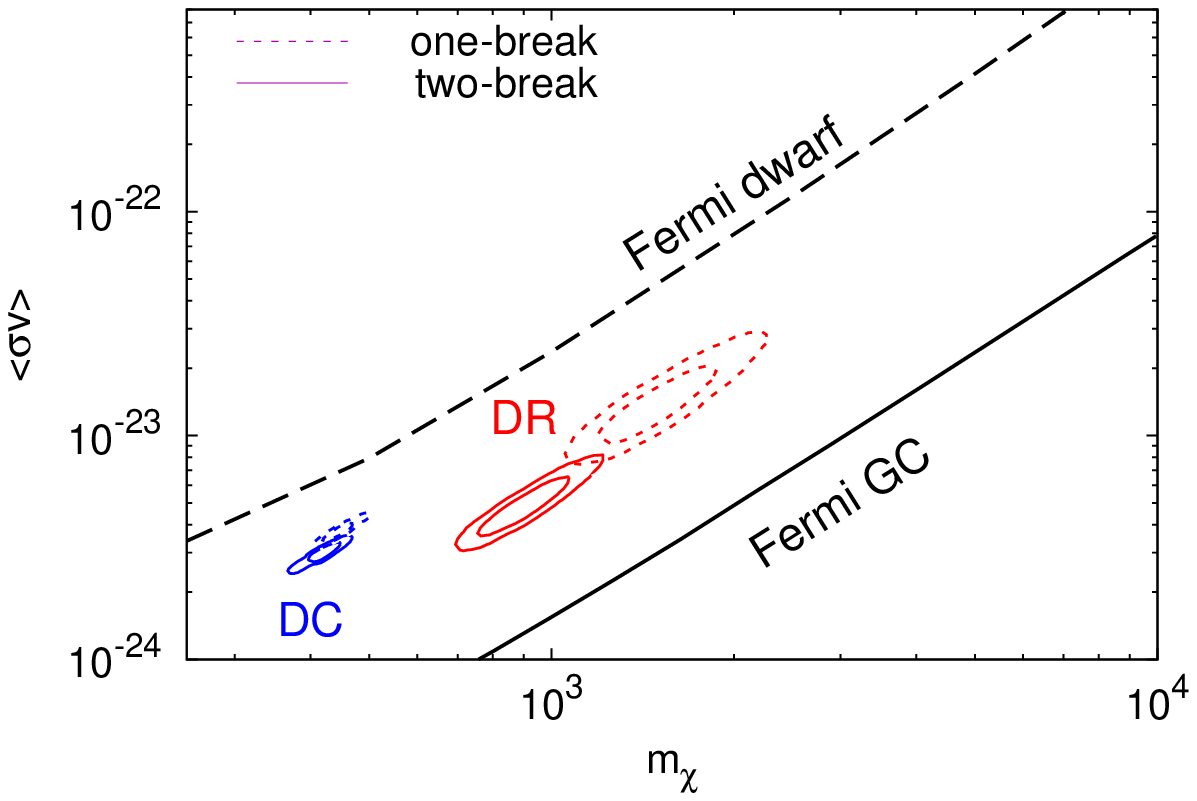}
  \includegraphics[width=0.48\textwidth]{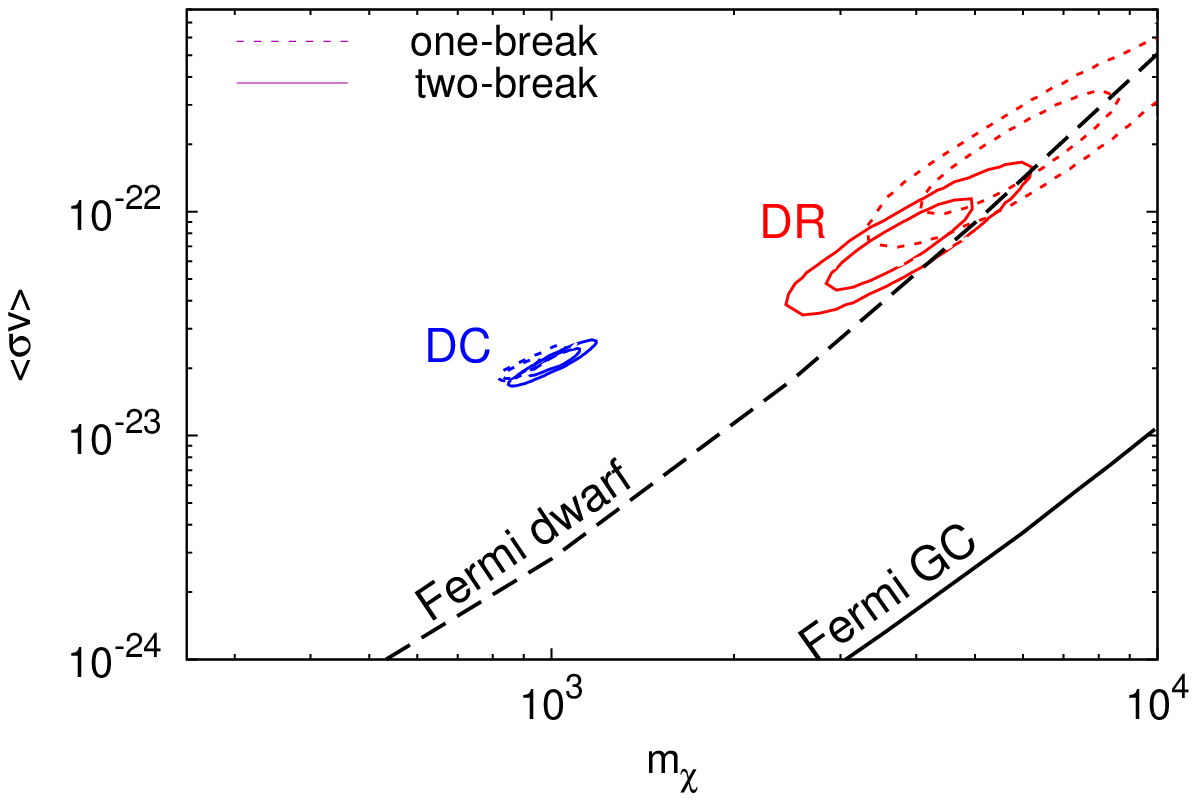}
  \caption{$1\sigma$ and $2\sigma$ confidence regions in $m_\chi-\langle
\sigma v\rangle$ plane, together with the exclusion lines from Fermi
$\gamma$-ray observations \cite{Ackermann:2013yva,Huang:2012yf}.
The left panel is for $\mu$ channel, while the right panel is for
$\tau$ channel. Blue contours are for DC scenario and red ones are for
DR scenario. The dashed regions are for the case with only one break
in the primary electron injection spectrum, while the solid regions are
for those with an additional break.}
  \label{fig:gamma}
\end{figure*}

Finally, it is well known that the DM annihilation scenario of the positron
excess is strongly constrained by the $\gamma$-ray observation.
The exclusion limit derived from $\gamma$-ray observations of the dwarf
galaxies by Fermi \cite{Ackermann:2013yva} and the Galactic center
\cite{Huang:2012yf} are shown in Fig. \ref{fig:gamma}. Similar to the
conclusion in \cite{Yuan:2013eja}, the constraints from the Galactic
center observation excludes all the DM scenarios. But this results suffer
from large uncertainty of the small scale DM density profile.
The constraints from the dwarf galaxies are much more solid. The DM
annihilation to $\tau$ channel shows tension with the $\gamma$-ray
observations. For $\mu$ channel, the current $\gamma$-ray data from
the dwarf galaxies can still not be able to exclude the required parameter
region to explain the lepton excess.

\section{Conclusions} \label{section_conclusions}

In this paper we give a global fitting to the AMS-02 new results of the
positron fraction, electron plus positron, electron and positron spectra to determine
the primary electron spectrum as well as the extra
$e^+e^-$ sources such as pulsars or DM. Two typical CR propagation
models, DR and DC models, are discussed.

We find that in order to fit the data an additional break (hardening) at
the primary electron spectrum at $\sim 60$ GeV is necessary. With such a
primary electron spectrum, both the pulsar scenario and the DM scenario
can give good fit to data, with $\chi^2$/d.o.f. close to or smaller than $1$.
The fittings are too worse to be acceptable without the additional
hardening of the primary spectrum. The best case without the additional
break is the DC scenario with pulsars as the extra source, which gives
$\chi^2$/d.o.f.$=374.2/264$ and corresponds to a $4.4\sigma$ deviation
from expectation.

In the two-break electron background case, both the pulsar and DM model 
can give good fittings to the AMS-02 lepton data. If DM annihilate into 
$\mu$ final state the fitting value of its mass is about $0.4-1.5$ TeV 
and the annihilation cross section is about $(3-10)\times 10^{-24}$ 
cm$^{3}$s$^{-1}$. For the $\tau$ final state the DM mass is about $1-7$ 
TeV and cross section is about $(2-20)\times 10^{-23}$ cm$^{3}$s$^{-1}$. 
It is interesting to note that the DM scenario can reproduce the potential 
drop of the positron fraction data at $\sim 300$ GeV with the best fitting 
mass values.

We further find that to fit the lepton data the DC propagation model is
more favorable than the DR model. This is because DR model will
induce a bump at the local interstellar spectrum as a consequence of
reacceleration. Such a bump is favored by the HEAO B/C data but is
not favored by the lepton spectra. Therefore it is very important for the
AMS-02 to give independent measurement of B/C down to sub-GeV so as to
determine the propagation model.

\acknowledgments
We thank Z.-H. Li, Z.-C. Tang, Z.-L. Weng and W.-W. Xu for helpful discussion.
This work is supported by the NSFC under Grant Nos. 11475191, 11135009, and
by the Strategic Priority Research Program
``The Emergence of Cosmological Structures'' of the Chinese
Academy of Sciences, under Grant No. XDB09000000.

\bibliography{ams}
\end{document}

%% file: tab/paraPSRnbk.tex
\begin{tabular}{>{$}l<{$}c*{2}{rpc}}
\hline
&\multirow{2}{*}{Prior Range}& \multicolumn{2}{c}{DR} & & \multicolumn{2}{c}{DC} & \\
\cline{3-4} \cline{6-7}
& & \multicolumn{1}{c}{Best} & \multicolumn{1}{c}{Mean} && \multicolumn{1}{c}{Best} & \multicolumn{1}{c}{Mean} & \\
\hline
\log(A_e\footnote{Post-propagated normalization flux of $e^-$ at $25\GeV$ in unit $\cm^{-2}\s^{-1}\sr^{-1}\MeV^{-1}$.}) & [-10.5, -7.5] & -8.812 & -8.813, 0.002 & & -8.897 & -8.897, 0.002 &\\
\gamma_1 & [1.0, 3.0] & 1.896 & 1.878, 0.025 & & 2.268 & 2.264, 0.02 &\\
\gamma_2 & [1.5, 4.0] & 2.874 & 2.872, 0.005 & & 2.725 & 2.726, 0.005 &\\
\log(R^e_\mathrm{br}/\MV) & [3.0, 6.0] & 3.697 & 3.68, 0.025 & & 3.812 & 3.808, 0.014 &\\
\log(A_\mathrm{psr}\footnote{Pre-propagated normalization of pulsar injection at $1\MeV$ in unit $\cm^{-3}\s^{-1}\MeV^{-1}$.}) & [-35.0, -20.0] & -25.99 & -25.91, 0.21 & & -25.31 & -25.35, 0.12 &\\
\alpha & [1.0, 2.4] & 1.749 & 1.763, 0.039 & & 1.859 & 1.851, 0.022 &\\
\log(R_\mathrm{c}/\MeV) & [4.0, 10.0] & 9.621 & 8.88, 0.7 & & 9.168 & 8.782, 0.77 &\\
c_{e^+} & [0.25, 4.0] & 2.63 & 2.599, 0.057 & & 1.204 & 1.216, 0.033 &\\
\phi/\MV & [100, 1500] & 1410.0 & 1402.0, 13.0 & & 562.1 & 565.9, 12.0 &\\
\hline
\end{tabular}

%% file: tab/paraMUnbk.tex
\begin{tabular}{>{$}l<{$}c*{2}{rpc}}
\hline
&\multirow{2}{*}{Prior Range}& \multicolumn{2}{c}{DR} & & \multicolumn{2}{c}{DC} & \\
\cline{3-4} \cline{6-7}
& & \multicolumn{1}{c}{Best} & \multicolumn{1}{c}{Mean} && \multicolumn{1}{c}{Best} & \multicolumn{1}{c}{Mean} & \\
\hline
\log(A_e\footnote{Post-propagated normalization flux of $e^-$ at $25\GeV$ in unit $\cm^{-2}\s^{-1}\sr^{-1}\MeV^{-1}$.}) & [-10.5, -7.5] & -8.801 & -8.801, 0.001 & & -8.872 & -8.872, 0.002 &\\
\gamma_1 & [1.0, 3.0] & 1.925 & 1.913, 0.018 & & 2.303 & 2.301, 0.037 &\\
\gamma_2 & [1.5, 4.0] & 2.9 & 2.9, 0.003 & & 2.79 & 2.789, 0.004 &\\
\log(R^e_\mathrm{br}/\MV) & [3.0, 6.0] & 3.706 & 3.692, 0.019 & & 3.702 & 3.704, 0.016 &\\
\log(m_\chi/\GeV) & [0.0, 7.0] & 3.179 & 3.178, 0.066 & & 2.654 & 2.65, 0.018 &\\
\log(\langle \sigma v\rangle\footnote{In unit $\cm^3\s^{-1}$}) & [-28.0, -18.0] & -22.85 & -22.86, 0.12 & & -23.42 & -23.43, 0.029 &\\
c_{e^+} & [0.25, 4.0] & 2.997 & 2.994, 0.019 & & 1.775 & 1.771, 0.014 &\\
\phi/\MV & [100, 1500] & 1488.0 & 1488.0, 6.7 & & 748.2 & 744.8, 7.9 &\\
\hline
\end{tabular}

%% file: tab/paraTAUnbk.tex
\begin{tabular}{>{$}l<{$}c*{2}{rpc}}
\hline
&\multirow{2}{*}{Prior Range}& \multicolumn{2}{c}{DR} & & \multicolumn{2}{c}{DC} & \\
\cline{3-4} \cline{6-7}
& & \multicolumn{1}{c}{Best} & \multicolumn{1}{c}{Mean} && \multicolumn{1}{c}{Best} & \multicolumn{1}{c}{Mean} & \\
\hline
\log(A_e\footnote{Post-propagated normalization flux of $e^-$ at $25\GeV$ in unit $\cm^{-2}\s^{-1}\sr^{-1}\MeV^{-1}$.}) & [-10.5, -7.5] & -8.803 & -8.803, 0.002 & & -8.883 & -8.882, 0.002 &\\
\gamma_1 & [1.0, 3.0] & 1.915 & 1.909, 0.019 & & 2.299 & 2.295, 0.031 &\\
\gamma_2 & [1.5, 4.0] & 2.895 & 2.895, 0.004 & & 2.757 & 2.758, 0.004 &\\
\log(R^e_\mathrm{br}/\MV) & [3.0, 6.0] & 3.698 & 3.691, 0.02 & & 3.728 & 3.722, 0.015 &\\
\log(m_\chi/\GeV) & [0.0, 7.0] & 3.78 & 3.786, 0.12 & & 2.936 & 2.954, 0.023 &\\
\log(\langle \sigma v\rangle\footnote{In unit $\cm^3\s^{-1}$}) & [-28.0, -18.0] & -21.72 & -21.71, 0.2 & & -22.72 & -22.69, 0.034 &\\
c_{e^+} & [0.25, 4.0] & 2.934 & 2.942, 0.026 & & 1.592 & 1.601, 0.016 &\\
\phi/\MV & [100, 1500] & 1472.0 & 1475.0, 8.8 & & 670.8 & 674.9, 8.7 &\\
\hline
\end{tabular}

%% file: tab/chi2tab.tex
\begin{tabular}{*{16}{c}}
\hline
\rule[-5.039999999999999pt]{0pt}{14.399999999999999pt} & & \multicolumn{6}{c}{two breaks} & \multicolumn{6}{c}{one break}\\
\cline{3-7}
\cline{10-15}
\rule[-8.399999999999999pt]{0pt}{24.0pt} &  & $\dfrac{\chi^2}{\mathrm{d.o.f.}}$ & $\chi^2$ & $\dfrac{e^+}{e^++e^-}$ & $e^++e^-$ & $e^-$ & $e^+$ & & $\dfrac{\chi^2}{\mathrm{d.o.f.}}$ & $\chi^2$ & $\dfrac{e^+}{e^++e^-}$ & $e^++e^-$ & $e^-$ & $e^+$ &\\[1ex]
\hline
\multirow{3}{*}{DR} & PSR & 1.1 & 287.1 & 48.57 & 99.02 & 60.03 & 79.45 &  & 2.75 & 725.2 & 132.7 & 271.8 & 211.7 & 109.0 &\\
 & $\mu$ & 1.12 & 293.4 & 40.7 & 113.8 & 61.0 & 77.95 &  & 3.01 & 797.9 & 242.4 & 245.5 & 206.9 & 103.1 &\\
 & $\tau$ & 1.11 & 291.0 & 42.62 & 106.4 & 63.38 & 78.6 &  & 2.94 & 779.9 & 213.9 & 252.2 & 209.9 & 103.9 &\\
\hline
\multirow{3}{*}{DC} & PSR & 0.411 & 107.6 & 51.58 & 17.33 & 17.62 & 21.07 &  & 1.42 & 374.2 & 89.04 & 121.3 & 119.1 & 44.78 &\\
 & $\mu$ & 1.27 & 334.8 & 116.1 & 90.8 & 35.97 & 91.93 &  & 3.95 & 1048.0 & 484.1 & 250.0 & 179.8 & 134.1 &\\
 & $\tau$ & 0.575 & 151.1 & 65.0 & 30.18 & 17.97 & 37.91 &  & 2.4 & 636.5 & 168.5 & 228.1 & 165.1 & 74.84 &\\
\hline
\end{tabular}

%% file: tab/paraPSRbk.tex
\begin{tabular}{>{$}l<{$}c*{2}{rpc}}
\hline
&\multirow{2}{*}{Prior Range}& \multicolumn{2}{c}{DR} & & \multicolumn{2}{c}{DC} & \\
\cline{3-4} \cline{6-7}
& & \multicolumn{1}{c}{Best} & \multicolumn{1}{c}{Mean} && \multicolumn{1}{c}{Best} & \multicolumn{1}{c}{Mean} & \\
\hline
\log(A_e\footnote{Post-propagated normalization flux of $e^-$ at $25\GeV$ in unit $\cm^{-2}\s^{-1}\sr^{-1}\MeV^{-1}$.}) & [-10.5, -7.5] & -8.813 & -8.812, 0.002 & & -8.896 & -8.897, 0.002 &\\
\gamma_1 & [1.0, 3.0] & 1.302 & 1.361, 0.054 & & 2.382 & 2.377, 0.022 &\\
\log(R^e_\mathrm{br}/\MV) & [3.0, 5.0] & 3.406 & 3.42, 0.02 & & 3.881 & 3.873, 0.02 &\\
\gamma_2 & [1.5, 4.0] & 2.976 & 2.972, 0.005 & & 2.836 & 2.829, 0.011 &\\
\log(R^e_\mathrm{br2}/\MV) & [4.0, 6.0] & 4.778 & 4.794, 0.028 & & 4.717 & 4.747, 0.044 &\\
\gamma_3 & [1.5, 4.0] & 2.668 & 2.656, 0.02 & & 2.586 & 2.571, 0.02 &\\
\log(A_\mathrm{psr}\footnote{Pre-propagated normalization of pulsar injection at $1\MeV$ in unit $\cm^{-3}\s^{-1}\MeV^{-1}$.}) & [-35.0, -20.0] & -28.88 & -28.71, 0.59 & & -26.8 & -26.77, 0.46 &\\
\alpha & [1.0, 2.4] & 1.185 & 1.221, 0.12 & & 1.564 & 1.569, 0.096 &\\
\log(R_\mathrm{c}/\MeV) & [4.0, 10.0] & 5.853 & 5.923, 0.22 & & 6.073 & 6.087, 0.23 &\\
c_{e^+} & [0.25, 4.0] & 3.029 & 3.02, 0.025 & & 1.53 & 1.512, 0.048 &\\
\phi/\MV & [100, 1500] & 1499.0 & 1495.0, 4.4 & & 672.8 & 667.3, 16.0 &\\
\hline
\end{tabular}

%% file: tab/paraMUbk.tex
\begin{tabular}{>{$}l<{$}c*{2}{rpc}}
\hline
&\multirow{2}{*}{Prior Range}& \multicolumn{2}{c}{DR} & & \multicolumn{2}{c}{DC} & \\
\cline{3-4} \cline{6-7}
& & \multicolumn{1}{c}{Best} & \multicolumn{1}{c}{Mean} && \multicolumn{1}{c}{Best} & \multicolumn{1}{c}{Mean} & \\
\hline
\log(A_e\footnote{Post-propagated normalization flux of $e^-$ at $25\GeV$ in unit $\cm^{-2}\s^{-1}\sr^{-1}\MeV^{-1}$.}) & [-10.5, -7.5] & -8.813 & -8.812, 0.001 & & -8.887 & -8.888, 0.002 &\\
\gamma_1 & [1.0, 3.0] & 1.501 & 1.53, 0.028 & & 2.427 & 2.427, 0.022 &\\
\log(R^e_\mathrm{br}/\MV) & [3.0, 5.0] & 3.473 & 3.48, 0.013 & & 3.849 & 3.847, 0.016 &\\
\gamma_2 & [1.5, 4.0] & 2.976 & 2.971, 0.005 & & 2.894 & 2.896, 0.01 &\\
\log(R^e_\mathrm{br2}/\MV) & [4.0, 6.0] & 4.787 & 4.803, 0.029 & & 4.712 & 4.713, 0.038 &\\
\gamma_3 & [1.5, 4.0] & 2.654 & 2.647, 0.021 & & 2.558 & 2.553, 0.023 &\\
\log(m_\chi/\GeV) & [0.0, 7.0] & 2.964 & 2.957, 0.049 & & 2.621 & 2.621, 0.02 &\\
\log(\langle \sigma v\rangle\footnote{In unit $\cm^3\s^{-1}$}) & [-28.0, -18.0] & -23.3 & -23.31, 0.085 & & -23.53 & -23.53, 0.032 &\\
c_{e^+} & [0.25, 4.0] & 3.053 & 3.049, 0.013 & & 1.855 & 1.858, 0.015 &\\
\phi/\MV & [100, 1500] & 1500.0 & 1498.0, 2.2 & & 781.8 & 784.8, 8.2 &\\
\hline
\end{tabular}

%% file: tab/paraTAUbk.tex
\begin{tabular}{>{$}l<{$}c*{2}{rpc}}
\hline
&\multirow{2}{*}{Prior Range}& \multicolumn{2}{c}{DR} & & \multicolumn{2}{c}{DC} & \\
\cline{3-4} \cline{6-7}
& & \multicolumn{1}{c}{Best} & \multicolumn{1}{c}{Mean} && \multicolumn{1}{c}{Best} & \multicolumn{1}{c}{Mean} & \\
\hline
\log(A_e\footnote{Post-propagated normalization flux of $e^-$ at $25\GeV$ in unit $\cm^{-2}\s^{-1}\sr^{-1}\MeV^{-1}$.}) & [-10.5, -7.5] & -8.813 & -8.812, 0.002 & & -8.891 & -8.891, 0.002 &\\
\gamma_1 & [1.0, 3.0] & 1.502 & 1.528, 0.025 & & 2.409 & 2.402, 0.021 &\\
\log(R^e_\mathrm{br}/\MV) & [3.0, 5.0] & 3.471 & 3.48, 0.012 & & 3.842 & 3.84, 0.015 &\\
\gamma_2 & [1.5, 4.0] & 2.972 & 2.969, 0.005 & & 2.855 & 2.856, 0.009 &\\
\log(R^e_\mathrm{br2}/\MV) & [4.0, 6.0] & 4.789 & 4.803, 0.029 & & 4.756 & 4.746, 0.04 &\\
\gamma_3 & [1.5, 4.0] & 2.656 & 2.651, 0.021 & & 2.548 & 2.555, 0.021 &\\
\log(m_\chi/\GeV) & [0.0, 7.0] & 3.59 & 3.581, 0.081 & & 3.003 & 3.006, 0.028 &\\
\log(\langle \sigma v\rangle\footnote{In unit $\cm^3\s^{-1}$}) & [-28.0, -18.0] & -22.13 & -22.14, 0.14 & & -22.68 & -22.67, 0.039 &\\
c_{e^+} & [0.25, 4.0] & 3.035 & 3.027, 0.015 & & 1.719 & 1.727, 0.018 &\\
\phi/\MV & [100, 1500] & 1500.0 & 1496.0, 3.1 & & 728.9 & 732.6, 9.3 &\\
\hline
\end{tabular}